\documentclass[useAMS,usenatbib]{mn2e}
\usepackage{epsfig} 
\usepackage{color} 

\def\mgii{Mg~{\sc ii}~} 
\def\mgiia{Mg~{\sc ii}$\lambda$2796~} 
\def\mgiib{Mg~{\sc ii}$\lambda$2803~} 
\def\mgiiab{Mg~{\sc ii}$\lambda\lambda$2796,2803~} 
\def\feii{Fe~{\sc ii}~} 

\def\oiiiab{O~{\sc iii}$\lambda\lambda$4959,5007~} 
\def\oiiia{O~{\sc iii}$\lambda$4959~} 
\def\oiiib{O~{\sc iii}$\lambda$5007~}

\def\hbeta{H$\beta$~}

\def\chisq{$\chi^{2}~$} 
\def\kms{km~s$^{-1}~$} 
\def\mbh{M$_{bh}$}
\def\aj{{AJ}}%
\def\araa{{ARA\&A}}%
\def\apj{{ApJ}}%
\def\aap{{A\&A}}%
\def\mnras{{MNRAS}}%
\def\pasp{{PASP}}%
\title[Spectra of microvariable
radio-quiet QSOs]{Probing spectral properties of radio-quiet quasars searched for
optical microvariability-II}
\author[Joshi R. et al. ]{Ravi Joshi$^{1}$\thanks{E-mail: ravi@aries.res.in (RJ); hum@aries.res.in (HC); wiitap@tcnj.edu (PJW); alok@aries.res.in (ACG), anand@iucaa.ernet.in (RS)},
Hum Chand$^1$, Paul J.\ Wiita$^{2}$, Alok C.\ Gupta$^{1}$, Raghunathan Srianand$^3$\\
$^{1}$Aryabhatta Research Institute of Observational Sciences (ARIES),
Manora Peak, Nainital $-$ 263129, India\\
$^{2}$Department of Physics, The College of New Jersey, P.O. Box 7718, Ewing, NJ 08628-0718, USA\\
$^{3}$ IUCAA, Postbag 4, Ganeshkhind, Pune 411 007, India }
\begin{document}
\date{Accepted -----. Received ------; in original form ----}

\pagerange{\pageref{firstpage}--\pageref{lastpage}} \pubyear{2011}

\maketitle

\label{firstpage}
\begin{abstract}

In the context of AGN unification scheme rapid variability
properties play an important role in understanding any intrinsic
differences between sources in different classes.  In this respect
any clue based on spectral properties will be very useful toward
understanding the mechanisms responsible for the origin of rapid
small scale optical variations, or microvariability. Here we have
added spectra of 46 radio-quiet quasars (RQQSOs) and Seyfert 1
galaxies to those of our previous sample of 37 such objects, all of
which had been previously searched for microvariability.  We
took new optical spectra of 33 objects and obtained 13 others from the
literature.  Their \hbeta and \mgii emission lines were carefully fit
to determine line widths (FWHM) as well as equivalent widths (EW) due
to the broad emission line components. The line widths were used to
estimate black hole masses and Eddington ratios, $\ell$. Both EW and
FWHM are anticorrelated with $\ell$.  Nearly all trends were in
agreement with our previous work, although the tendency for
sources exhibiting microvariability to be of lower luminosity was
not confirmed.  Most importantly, this whole sample of EW
distributions provides no evidence for the hypothesis that a weak jet
component in the radio quiet AGNs is responsible for their
microvariability.

\end{abstract}
%% each reference.
\section{Introduction}

{ 
Rapid small scale optical variations, or intra-night
microvariability is a well known characteristic of Active Galactic
Nuclei (AGNs) but the processes causing the bulk of these
microvariations is still a matter of debate.  The variability
mechanisms in the radio-loud objects are widely believed to be
connected to conditions in the relativistic jet. However, it is still
unclear if in the radio-quiet objects the nature of intra-night
variability is different, or whether the faint, variable jet is also a
dominant component for them.  Czerny et al.\ (2008) have tried to
understand the microvariation mechanism within the framework of
following scenarios: (i) fluctuations from an accretion disk (e.g.,
Mangalam \& Wiita 1993); (ii) irradiation of an accretion disc by a
variable X-ray flux (e.g., Rokaki, Collin-Souffrin \& Magnan 1993;
Gaskell 2006); and (iii) the presence of modestly misaligned jets in
radio-quiet quasars or a ``blazar component'' (e.g., Gopal-Krishna et
al.\ 2003). They concluded that the blazar component model is the most
promising to give rise to intra-night optical variability.

In this blazar component scenario, spectral properties of the sources
can play a crucial role in constraining the models further.  For
instance, if blazar components are dominating the variability of
RQQSOs, then, due to the increase in the continuum produced by the
jets one expects smaller equivalent widths (EW) of prominent emission
lines such as \hbeta and \mgii for sources with microvaraibility as
compared to their average value in a sample including non-variable
sources.  Recently, in Chand, Wiita \& Gupta (2010; hereafter Paper
I), we have worked toward this goal by exploiting the optical spectra
available from the Sloan Digital Sky Survey, Data Release 7 (SDSS
DR-7; Abazajian et al.\ 2009).  We carried out careful spectral
modeling of the \hbeta and \mgii emission line regions for RQQSO and
Seyfert 1 samples already searched for microvariability (hereafter
referred to collectively as RQQSOs).  In Paper I, we first
investigated any effect of key spectral parameters (e.g., EW and FWHM)
on the microvariability of RQQSOs. Second, we estimated other relevant
AGN parameters such as the black hole mass, \mbh ~and Eddington
ratios. To do this, their \hbeta and \mgii emission lines were
carefully fit to determine line widths (FWHM) as well as equivalent
widths (EW) due to the broad emission line components.  The line
widths were used to estimate black hole masses and Eddington ratios. The EW distributions did not provide evidence for the
hypothesis that a weak jet component in the RQQSOs is responsible for
their microvariability, and perhaps instead it may indicate that
variations involving the accretion disc (e.g., Wiita 2006) are
important for them.  We also concluded that there may be a weak
negative correlation between \hbeta EW and $\ell$, but there is a significant one between the \mgii EW and
$\ell$. In addition, we noted that there is a tendency for sources
with detectable optical microvariability to have somewhat lower
luminosities than those with no such detections (see Fig.\ 8 of Paper
I) .\par

All the above tentative conclusions from our Paper I were certainly
interesting; however, as we stressed there, it is very important that
they should be tested by examining larger samples. The importance of a
larger sample for such investigations is evident from our recent
finding on the fraction of broad absorption line QSOs showing
microvariability (Joshi et al.\ 2011).  That fraction appeared to be
50 per cent based on only 6 sources, but using a larger sample of 23
sources, Joshi et al. (2011) found it to be around ten percent,
similar to RQQSOs (Gupta \& Joshi 2005).

Therefore, given the important implications of our above results,
based on spectral analysis of 37 sources, for the origin of RQQSO
microvariability, and hence to processes on the accretion disc itself,
it becomes important to carry out spectral analyses of as many sources
searched for microvariability as possible.  The obvious source for
such a sample would be the remainder of the objects from the total of
117 sources in the compilation of Carini et al. (2007) for which SDSS
DR-7 spectra are not available. This forms the main motivation of this
paper, in which we more than double the sample size by taking new
spectra and gathering others from the literature that we then
analyze. Results from this larger sample can test the validity of
results found using our previous modest sample in Paper I.\par

The paper is organized as follows. Section 2 describes the data sample
and selection criteria. Section 3 describes the observation and data
reduction while Section 4 gives details of our spectral fitting
procedure. In Section 5 we focus on BH mass measurements and in
Section 6 we give estimates of Eddington ratios and of BH growth
times.  Section 7 gives a discussion and conclusions. Throughout, we
have used flat cosmology with $H_{\rm 0}$=70 km\,s$^{-1}$\,Mpc$^{-1}$,
 $\Omega_{\rm m}$=0.3} and $\Omega_{\rm \Lambda}$=0.7.

%%%%%
\begin{table*}
 \centering
 \begin{minipage}{140mm}
\caption{ New sample of radio-quiet QSOs and Seyfert
galaxies from the compilation of Carini et al.\ (2007)}
\label{lab:tabsamp}
\begin{tabular}{@{}ccc ccc cc l@{}} 
\hline 
\multicolumn{1}{c}{QSO Name\footnote{Object name in J{2000}}}  
& \multicolumn{1}{c}{$\alpha_{2000}$} 
& \multicolumn{1}{c}{$\delta_{2000}$} 
& \multicolumn{1}{c}{$z_{em}$ \footnote{Redshift as determined from the 
peak of  \oiiib emission line. \\ 
$\star$ Apparent  magnitudes (m$_{B}$ or m$_{V}$)  and absolute magnitudes (M$_{B}$)  are taken from Carini et al.\ (2007).}}   
&{m$_{(band)}^{\star}$ }
& {M$_{B}^{\star}$} 
&  {Variable?} 
& {Class} 
& \multicolumn{1}{c}{Spectra \footnote {NED spectral references: $^{\it 1}$Moustakas 
    \& Kennicutt (2006); $^{\it 2,3,4}$Boroson \& Green (1992) ; $^{\it 5}$Kim et al. \ (1995).}}
\\
(1)&(2)&(3)&(4)&(5)&(6)&(7)&(8)&(9)\\
%$^{\spadesuit}$ \\ 
\hline 
\\
J000619.5$+$201210            &  00$^h$    06$^m$    19.5$^s$ &   $+$20$^\circ$   12$^\prime$   10$^{\prime\prime}$ & 0.025&  13.75$_{(B)}$ &   $$-$$22.14& N  &SY1.2 & IGO      \\    
 J002913.6$+$131603            &  00$^h$    29$^m$    13.6$^s$ &   $+$13$^\circ$   16$^\prime$   03$^{\prime\prime}$ & 0.142&  16.30$_{(V)}$ &   $$-$$24.70& Y  &SY1   & IGO        \\
J004547.3$+$041024            &  00$^h$    45$^m$    47.3$^s$ &   $+$04$^\circ$   10$^\prime$   24$^{\prime\prime}$ & 0.385&  15.88$_{(B)}$ &   $$-$$26.00& N &BALQSO & IGO        \\
J005334.9$+$124136            &  00$^h$    53$^m$    34.9$^s$ &   $+$12$^\circ$   41$^\prime$   36$^{\prime\prime}$ & 0.058&  14.39$_{(B)}$ &   $$-$$24.43& N  &SY1   & IGO        \\
J005452.1$+$252538            &  00$^h$    54$^m$    52.1$^s$ &   $+$25$^\circ$   25$^\prime$   38$^{\prime\prime}$ & 0.154&  15.42$_{(B)}$ &   $$-$$24.40& N  &SY1   & IGO        \\
J011354.5$+$390744            &  01$^h$    13$^m$    54.5$^s$ &   $+$39$^\circ$   07$^\prime$   44$^{\prime\prime}$ & 0.234&  16.70$_{(B)}$ &   $$-$$24.10& N  &SY1   & IGO        \\
J012240.6$+$231015            &  01$^h$    22$^m$    40.6$^s$ &   $+$23$^\circ$   10$^\prime$   15$^{\prime\prime}$ & 0.052&  15.41$_{(B)}$ &   $$-$$22.12& N  &SY1   & IGO        \\
J051611.4$-$000859            &  05$^h$    16$^m$    11.4$^s$ &   $-$00$^\circ$   08$^\prime$   59$^{\prime\prime}$ & 0.032&  14.10$_{(B)}$ &   $$-$$25.20& Y  &SY1   & IGO        \\
J051633.4$-$002713            &  05$^h$    16$^m$    33.4$^s$ &   $-$00$^\circ$   27$^\prime$   13$^{\prime\prime}$ & 0.292&  16.26$_{(B)}$ &   $$-$$25.10& N  &SY1.2 & IGO        \\
J055453.6$+$462622            &  05$^h$    54$^m$    53.6$^s$ &   $+$46$^\circ$   26$^\prime$   22$^{\prime\prime}$ & 0.020&  15.00$_{(B)}$ &   $$-$$22.60& N  &SY1   & IGO        \\
J071415.1$+$454156            &  07$^h$    14$^m$    15.1$^s$ &   $+$45$^\circ$   41$^\prime$   56$^{\prime\prime}$ & 0.055&  14.90$_{(B)}$ &   $$-$$29.60& N  &SY1.5 & IGO        \\
J073657.0$+$584613            &  07$^h$    36$^m$    57.0$^s$ &   $+$58$^\circ$   46$^\prime$   13$^{\prime\prime}$ & 0.039&  15.29$_{(B)}$ &   $$-$$21.80& N  &SY1.5 & IGO        \\
J084742.4$+$344504            &  08$^h$    47$^m$    42.4$^s$ &   $+$34$^\circ$   45$^\prime$   04$^{\prime\prime}$ & 0.064&  14.00$_{(B)}$ &   $$-$$23.95& Y  &SY1   & IGO       \\
J092512.9$+$521711            &  09$^h$    25$^m$    12.9$^s$ &   $+$52$^\circ$   17$^\prime$   11$^{\prime\prime}$ & 0.035&  15.62$_{(B)}$ &   $$-$$21.00& Y  &SY1   & SDSS      \\
J092603.3$+$124404            &  09$^h$    26$^m$    03.3$^s$ &   $+$12$^\circ$   44$^\prime$   04$^{\prime\prime}$ & 0.029&  14.93$_{(B)}$ &   $$-$$21.27& N  &SY1.2 & SDSS       \\
J095652.4$+$411522            &  09$^h$    56$^m$    52.4$^s$ &   $+$41$^\circ$   15$^\prime$   22$^{\prime\prime}$ & 0.234&  15.05$_{(B)}$ &   $$-$$25.73& N  &SY1   & IGO        \\
J101420.7$-$041840            &  10$^h$    14$^m$    20.7$^s$ &   $-$04$^\circ$   18$^\prime$   40$^{\prime\prime}$ & 0.058&  15.49$_{(B)}$ &   $$-$$22.22& N  &SY1   & IGO        \\
J105143.9$+$335927            &  10$^h$    51$^m$    43.9$^s$ &   $+$33$^\circ$   59$^\prime$   27$^{\prime\prime}$ & 0.167&  15.81$_{(B)}$ &   $$-$$24.19& N  &SY1   & SDSS      \\
J110631.8$-$005252            &  11$^h$    06$^m$    31.8$^s$ &   $-$00$^\circ$   52$^\prime$   52$^{\prime\prime}$ & 0.423&  16.02$_{(B)}$ &   $$-$$25.70& Y  &QSO   & IGO        \\
J111908.7$+$211918            &  11$^h$    19$^m$    08.7$^s$ &   $+$21$^\circ$   19$^\prime$   18$^{\prime\prime}$ & 0.176&  15.17$_{(B)}$ &   $$-$$24.96& Y  &SY1   & IGO         \\
J112147.1$+$114418            &  11$^h$    21$^m$    47.1$^s$ &   $+$11$^\circ$   44$^\prime$   18$^{\prime\prime}$ & 0.050&  14.65$_{(B)}$ &   $$-$$22.69& N  &SY1.2 & IGO        \\
J112302.3$-$273004            &  11$^h$    23$^m$    02.3$^s$ &   $-$27$^\circ$   30$^\prime$   04$^{\prime\prime}$ & 0.389&  16.80$_{(V)}$ &   $$-$$25.20& N  &QSO   & IGO         \\
J112439.2$+$420145            &  11$^h$    24$^m$    39.2$^s$ &   $+$42$^\circ$   01$^\prime$   45$^{\prime\prime}$ & 0.225&  16.02$_{(B)}$ &   $$-$$24.71& N  &SY 1  & SDSS       \\ 
J112731.9$-$304446            &  11$^h$    27$^m$    31.9$^s$ &   $-$30$^\circ$   44$^\prime$   46$^{\prime\prime}$ & 0.673&  16.30$_{(V)}$ &   $$-$$27.00& N  &QSO   & IGO        \\  
J115349.3$+$112830            &  11$^h$    53$^m$    49.3$^s$ &   $+$11$^\circ$   28$^\prime$   30$^{\prime\prime}$ & 0.176&  15.51$_{(B)}$ &   $$-$$24.61& N  &SY 1  & SDSS       \\
J120309.6$+$443153            &  12$^h$    03$^m$    09.6$^s$ &   $+$44$^\circ$   31$^\prime$   53$^{\prime\prime}$ & 0.002&  13.74$_{(B)}$ &   $$-$$17.40& N  &SY1.5 & NED$^{\bf \it 1}$     \\%2006ApJS..164...81       \\
J121032.6$+$392421            &  12$^h$    10$^m$    32.6$^s$ &   $+$39$^\circ$   24$^\prime$   21$^{\prime\prime}$ & 0.003&  11.50$_{(B)}$ &   $$-$$20.60& N  &SY1.5 & IGO        \\
J125048.3$+$395139            &  12$^h$    50$^m$    48.3$^s$ &   $+$39$^\circ$   51$^\prime$   39$^{\prime\prime}$ & 1.032&  16.06$_{(V)}$ &   $$-$$27.86& N  &QSO   & SDSS       \\
J125948.8$+$342323            &  12$^h$    59$^m$    48.8$^s$ &   $+$34$^\circ$   23$^\prime$   23$^{\prime\prime}$ & 1.375&  16.79$_{(B)}$ &   $$-$$28.00& N  &QSO   & SDSS       \\
J132349.5$+$654148            &  13$^h$    23$^m$    49.5$^s$ &   $+$65$^\circ$   41$^\prime$   48$^{\prime\prime}$ & 0.168&  15.86$_{(B)}$ &   $$-$$24.23& N  &SY1   & IGO        \\
J135458.7$+$005211            &  13$^h$    54$^m$    58.7$^s$ &   $+$00$^\circ$   52$^\prime$   11$^{\prime\prime}$ & 1.127&  16.00$_{(V)}$ &   $$-$$28.06& N  &QSO   & IGO        \\
J140516.2$+$255534            &  14$^h$    05$^m$    16.2$^s$ &   $+$25$^\circ$   55$^\prime$   34$^{\prime\prime}$ & 0.164&  15.57$_{(B)}$ &   $$-$$24.46& N  &SY1   & IGO         \\
J141348.3$+$440014            &  14$^h$    13$^m$    48.3$^s$ &   $+$44$^\circ$   00$^\prime$   14$^{\prime\prime}$ & 0.089&  14.99$_{(B)}$ &   $$-$$23.68& N  &SY1   & IGO         \\
J141700.7$+$445606            &  14$^h$    17$^m$    00.7$^s$ &   $+$44$^\circ$   56$^\prime$   06$^{\prime\prime}$ & 0.113&  15.74$_{(B)}$ &   $$...$$   & N  &SY1   & SDSS        \\
J142906.6$+$011706            &  14$^h$    29$^m$    06.6$^s$ &   $+$01$^\circ$   17$^\prime$   06$^{\prime\prime}$ & 0.086&  15.05$_{(B)}$ &   $$-$$23.51& N  &SY1   & IGO         \\
J144207.4$+$352623            &  14$^h$    42$^m$    07.4$^s$ &   $+$35$^\circ$   26$^\prime$   23$^{\prime\prime}$ & 0.079&  15.00$_{(B)}$ &   $$-$$23.32& N  &SY1   & IGO         \\
J153638.3$+$543333            &  15$^h$    36$^m$    38.3$^s$ &   $+$54$^\circ$   33$^\prime$   33$^{\prime\prime}$ & 0.038&  15.31$_{(B)}$ &   $$-$$21.48& N  &SY1   & IGO         \\
J155202.3$+$201402            &  15$^h$    52$^m$    02.3$^s$ &   $+$20$^\circ$   14$^\prime$   02$^{\prime\prime}$ & 0.251&  17.40$_{(V)}$ &   $$-$$24.60& N  &QSO   & IGO           \\
J162011.3$+$172428            &  16$^h$    20$^m$    11.3$^s$ &   $+$17$^\circ$   24$^\prime$   28$^{\prime\prime}$ & 0.112&  15.53$_{(B)}$ &   $$...$$   & N  &SY1   & IGO           \\
J170124.8$+$514920            &  17$^h$    01$^m$    24.8$^s$ &   $+$51$^\circ$   49$^\prime$   20$^{\prime\prime}$ & 0.292&  15.43$_{(B)}$ &   $$-$$25.78& Y&BALQSO  & NED$^{\bf \it 2}$     \\%1992ApJS...80..109          \\
J175116.6$+$504539            &  17$^h$    51$^m$    16.6$^s$ &   $+$50$^\circ$   45$^\prime$   39$^{\prime\prime}$ & 0.299&  15.80$_{(B)}$ &   $$-$$25.60& Y  &QSO   & IGO           \\
J211452.6$+$060742            &  21$^h$    14$^m$    52.6$^s$ &   $+$06$^\circ$   07$^\prime$   42$^{\prime\prime}$ & 0.466&  15.52$_{(B)}$ &   $$-$$26.70& N  &QSO   & IGO           \\
J213227.8$+$100819            &  21$^h$    32$^m$    27.8$^s$ &   $+$10$^\circ$   08$^\prime$   19$^{\prime\prime}$ & 0.062&  14.62$_{(B)}$ &   $$-$$23.20& N  &QSO   & NED$^{\bf \it 3}$     \\%1992ApJS...80..109          \\
J221712.2$+$141421            &  22$^h$    17$^m$    12.2$^s$ &   $+$14$^\circ$   14$^\prime$   21$^{\prime\prime}$ & 0.065&  14.98$_{(B)}$ &   $$-$$23.04& N  &QSO   & NED$^{\bf \it 4}$     \\%1992ApJS...80..109          \\
J230315.6$+$085226            &  23$^h$    03$^m$    15.6$^s$ &   $+$08$^\circ$   52$^\prime$   26$^{\prime\prime}$ & 0.016&  13.00$_{(B)}$ &   $$-$$22.10& Y  &SY1.2 & NED$^{\bf \it 5}$     \\%1995ApJS...98..129          \\
J230702.9$+$043257            &  23$^h$    07$^m$    02.9$^s$ &   $+$04$^\circ$   32$^\prime$   57$^{\prime\prime}$ & 0.046&  15.44$_{(B)}$ &   $$-$$21.57& N  &SY1   & IGO           \\
\hline   
\end{tabular}                                                           
\end{minipage}                                                          
\end{table*}

%%%%%%%%%%%%%%%%SECTION 2 %%%%%%%%%%%%%%%%%%%%%%%%%%%%%%%%%%%%%%%%%%%%%%

\section{Data sample and selection criteria}

{ 

From the compilation of 117 radio-quiet AGNs in Carini et al.\ (2007),
that were often extensively searched for microvariability, Paper I
analyzed 33 sources (in their total sample of 37 sources), for which
SDSS DR7 spectra were available.  Here we examined the rest of the the
84 sources, and using their redshifts found that among them 25 sources
are such that neither their \hbeta nor \mgii lines fall in the optical
spectral range as they have redshifts outside the following
  respective ranges: $z \leq 0.65; 0.43 \leq z \leq 1.86$. For the
remaining 59 sources we found \hbeta and \mgii emission lines fall in
the spectral range of 3800\AA-8300\AA \ \ for 50 and 9 sources,
respectively.

We have searched for the optical spectra of these 59 sources in SDSS
DR8 (Aihara et al.\ 2011) and as well as in the NED data base.
We could find SDSS DR8 spectra for 8 sources and NED spectra with
desired quality (good S/N and spectral coverage that cover \mgii or
\hbeta line) for 5 sources as listed in Table~\ref{lab:tabsamp}.

We planned the observations of the rest of the 46 sources, using the
IUCAA Faint Object Spectrograph (IFOSC) mounted on the 2-m
telescope at IUCAA Girawali Observatory (IGO), near Pune, India.  With
IGO we could obtain the desired spectra for 33 sources. Among the
remaining 13 sources: 6 were not visible from IGO, having very
  negative declinations; 4 (J001555.1$+$023024, J012017.2$+$213346,
  J220311.5$-$180143 and J194240.6$-$101925) could not be observed
  either due to non-visibility or lack of enough observing time; 1,
  namely J103206.2$+$324015, was classified as a white dwarf in NED;
  and the remaining 2, J082740.2$+$094208 and J154559.1$+$270630,
though observed, were dropped from our analysis as they showed very
poor \hbeta lines in their spectra.

So finally we are left with a sample of 46 sources (33 new
observations, 8 from SDSS DR8 and 5 from NED); among them, 42 spectra
cover \hbeta lines while 4 spectra cover \mgii doublet lines. The
source information is listed in Table~\ref{lab:tabsamp}. The first
four columns give source name, RA(J2000), DEC(J2000) and emission
redshift (z$_{em}$), while the fifth column provides the apparent
magnitude in either B or V band. The sixth column provides the
absolute B magnitudes and the seventh column indicate microvariability
detection status; the eighth column gives a source classification from
among classes such as QSO, BALQSO (Broad Absorption Line QSO), or
  Seyfert (Sy) galaxy type. In the last column we have given the
spectral resource information.

%%%%%%%%%%%%%%%%%%%SECTION 3 %%%%%%%%%%%%%%%%%%
\section{Observations and Data Reduction}
\label{sec:analysis}
%%\subsection{Spectral Observations}

The new spectra were obtained for 33 sources. The observations were
carried out using the IFOSC mounted on the 2 meter IGO telescope.  We
took long-slit spectra covering the wavelength range 3800 - 6840~{\AA}
using the GR7 grism and 5800 - 8300~{\AA} using the GR8 grism of the
IFOSC to cover the \mgii and \hbeta lines, respectively, as noted in
the third column of the observing log in Table~\ref{lab:logtable}.
%, listed in last column of Table~\ref{lab:tabsamp}.  
A slit width of either 1.$^{\prime \prime}$0 or 1.$^{\prime \prime}$5 was used, as listed in
the  fourth column of Table~\ref{lab:logtable}.  Typical seeing during 
our observations was around 1.$^{\prime \prime}$2 to 1.$^{\prime \prime}$3.   
The raw CCD frames were
cleaned using standard IRAF{\footnote {IRAF is distributed by the
National Optical Astronomy Observatories, which are operated by
the Association of Universities for Research in Astronomy, Inc.,
under cooperative agreement with the National Science
Foundation.}}  procedures. We used halogen flats for flat fielding
the frames.  Since at $\lambda>7000${\AA} simple flat fielding does
not remove the fringes, therefore in case of grism GR8 the QSO was
moved along the slit for two different exposures at the same position
angle. We subsequently removed fringing by subtracting one frame from
the other taken on the same night.  The same procedure was applied to
standard stars as well. 
We then extracted the one dimensional spectrum from individual frames
using the IRAF task ``doslit''. Wavelength calibration of the spectra
was performed using Helium-Neon lamps. 
The spectrophotometric flux calibration was done using standard stars 
and assuming a mean extinction for the IGO site. In cases when multiple exposures
were needed we coadded  the flux with {${1}/{\sigma_i^2}$} weightage, where ${\sigma_i}$ 
is the error on the individual pixel.
The error spectrum was computed taking into account proper
error propagation during the combining process.  
The typical median signal-to-noise (SNR) over the spectral fitting range of our sample  vary from 20-150 per pixel, as listed in 
last column of Table~\ref{lab:logtable}. Our spectral range also covers atmospheric 
absorption line regions but in all cases our fitting regions fall either buleward or redward 
of them, so we need not  correct for any blending due to them in our analyses.

\section{Line analysis through simultaneous spectral fitting}
\label{sec:aalysis}

The \hbeta and \mgii emission lines were carefully modelled by using
the fitting procedure we used in Paper I.  The spectra were first
corrected for Galactic extinction using the extinction map of Schlegel
et al.\ (1998) and the reddening curve of Fitzpatrick (1999). Then
they were transformed into the rest frame using the redshift as
determined from the peak of \oiiib emission line.  Limited by the
complications needed to fit the continuum and the \feii
emission, {\it viz.}  (i) there are essentially no emission-line free
regions where the continuum can be determined (Vanden Berk et
al.\ 2001); (ii) the prominence of \feii features and their blending
with the \hbeta and \mgii lines; (iii) the \hbeta line is highly
blended with the \oiiiab lines, we have opted to carry out
simultaneous fits\footnote{To carry out the simultaneous fit we have
  used the \textsc{MPFIT} package for nonlinear fitting, written in
  \textsc{Interactive Data Language} routines.  MPFIT is kindly
  provided by Craig B. Markwardt and is available at
  http://cow.physics.wisc.edu/\~{}craigm/idl/.} of continuum, \feii
emission, \hbeta and \mgii and all other metal emission lines present
in the spectra. For this purpose we adopted the procedure as described
in detail in Paper I, which in brief is as follows.

%%%
\begin{table}
 \centering
 \begin{minipage}{140mm}
\caption{ IGO Observation log.}
\label{lab:logtable}
\begin{tabular}{@{}cc cc r@{}} 
\hline 
\multicolumn{1}{c}{QSO Name}
& \multicolumn{1}{c}{Date of Obs} 
& \multicolumn{1}{c}{Grism} 
&{slit width} 
&{SNR} 
\\
& & & & \\
%$^{\spadesuit}$ \\ 
\hline 
\\   
J000619.5$+$201210 &2009-12-13   &IFOSC7 &    1.5$''$ &   82 \\
J002913.6$+$131603 &2009-12-13   &IFOSC7 &    1.0$''$ &   73 \\
J004547.3$+$041024 &2009-12-12   &IFOSC8 &    1.5$''$ &   38 \\
J005334.9$+$124136 &2009-12-13   &IFOSC7 &    1.0$''$ &   80 \\
J005452.1$+$252538 &2009-12-12   &IFOSC7 &    1.5$''$ &   40 \\
J011354.5$+$390744 &2011-01-06   &IFOSC7 &    1.5$''$ &   19 \\
J012240.6$+$231015 &2011-02-05   &IFOSC7 &    1.5$''$ &   47 \\
J051611.4$-$000859 &2011-01-06   &IFOSC7 &    1.5$''$ &  127 \\ 
J051633.4$-$002713 &2009-12-12   &IFOSC8 &    1.5$''$ &   43 \\ 
J055453.6$+$462622 &2009-12-13   &IFOSC7 &    1.0$''$ &   47 \\ 
J071415.1$+$454156 &2011-02-04   &IFOSC7 &    1.0$''$ &   54 \\ 
J073657.0$+$584613 &2011-02-05   &IFOSC7 &    1.0$''$ &   63 \\ 
J084742.4$+$344504 &2009-12-12   &IFOSC7 &    1.5$''$ &   92 \\ 
J095652.4$+$411522 &2009-12-13   &IFOSC7 &    1.0$''$ &   44 \\ 
J101420.7$-$041840 &2011-02-05   &IFOSC7 &    1.0$''$ &   67 \\
J110631.8$-$005252 &2011-04-01   &IFOSC8 &    1.5$''$ &   39 \\ 
J111908.7$+$211918 &2010-04-18   &IFOSC7 &    1.0$''$ &   32 \\ 
J112147.1$+$114418 &2011-02-04   &IFOSC7 &    1.0$''$ &   41 \\ 
J112302.3$-$273004 &2011-04-01   &IFOSC8 &    1.5$''$ &   73 \\ 
J112731.9$-$304446 &2010-04-19   &IFOSC7 &    1.0$''$ &   42\\
J121032.6$+$392421 &2010-04-18   &IFOSC7 &    1.0$''$ &   21 \\ 
J132349.5$+$654148 &2011-02-04   &IFOSC7 &    1.0$''$ &  150 \\
J135458.7$+$005211 &2011-03-30   &IFOSC7 &    1.5$''$ &   37 \\
J140516.2$+$255534 &2011-02-04   &IFOSC7 &    1.0$''$ &   51 \\ 
J141348.3$+$440014 &2010-04-17   &IFOSC7 &    1.0$''$ &   40 \\ 
J142906.6$+$011706 &2010-04-18   &IFOSC7 &    1.0$''$ &   31 \\ 
J144207.4$+$352623 &2010-04-17   &IFOSC7 &    1.0$''$ &   50 \\ 
J153638.3$+$543333 &2010-04-17   &IFOSC7 &    1.0$''$ &   37 \\ 
J155202.3$+$201402 &2011-02-05   &IFOSC7 &    1.0$''$ &   23 \\ 
J162011.3$+$172428 &2010-04-18   &IFOSC7 &    1.0$''$ &   23 \\ 
J175116.6$+$504539 &2011-03-30   &IFOSC7 &    1.5$''$ &   55 \\ 
J211452.6$+$060742 &2009-12-12   &IFOSC8 &    1.5$''$ &   38 \\ 
J230702.9$+$043257 &2009-12-14   &IFOSC7 &    1.0$''$ &   20 \\

     \hline                                                                                                                                                                                
\end{tabular}                                                           
\end{minipage}                                                          
\end{table}

%%%%%%%%FIGURE 
\begin{figure*}
%\rotate
 \epsfig{figure=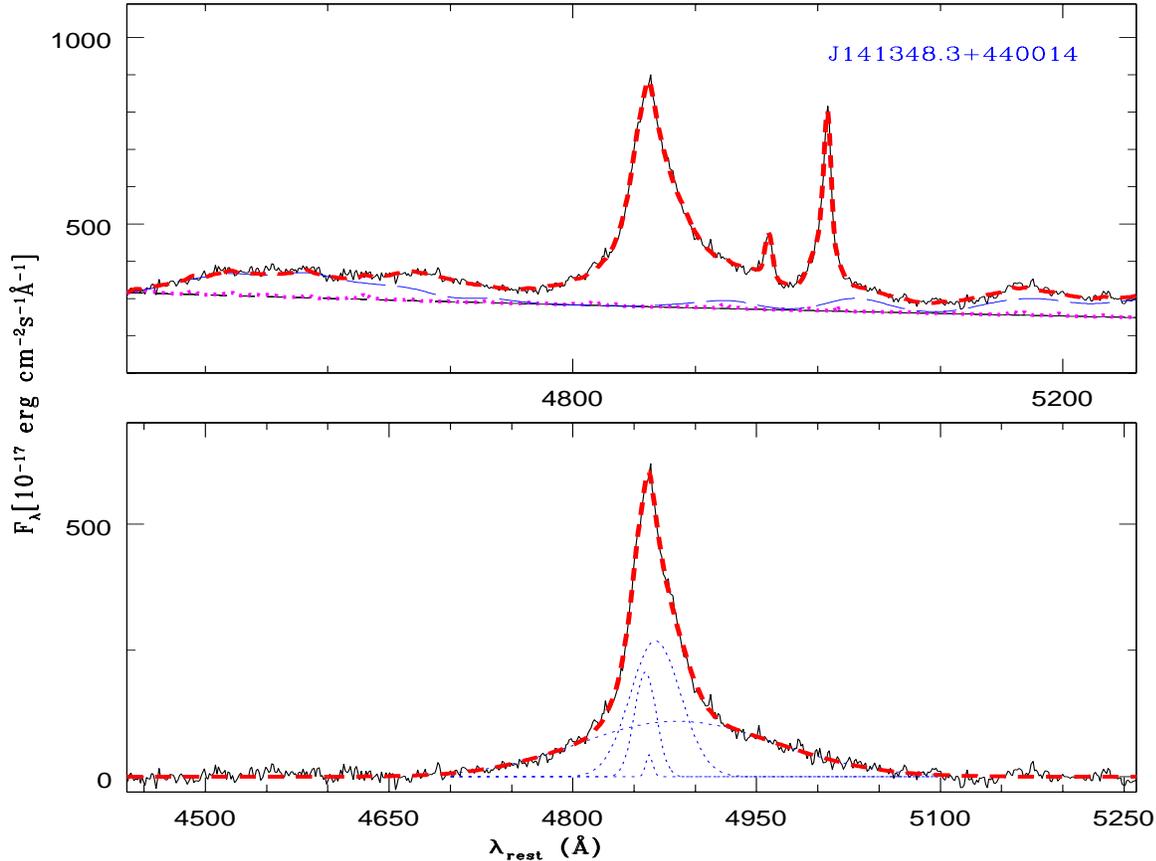,height=12.5cm,width=16.cm,angle=0} 
 \caption{The best fit to the \hbeta emission line of the 
IGO spectra of the QSO J141348.3$+$440014.  Upper panel:
complete spectrum fit (thick dashed/red) and components of the fit:
power law continuum (thin short-dashed/black line), broad \feii\ ({thin long-dashed}/blue),
narrow \feii\ (dotted/magenta)  lines. Lower panel:
continuum, \feii\ and metal line subtracted spectrum (solid/black)
  with the best fit total \hbeta profile (thick dashed/red) and
  \hbeta\ components (dotted/blue) lines.  Note that the entire fit is
  performed simultaneously (not first continuum subtraction then
  \hbeta fit) but these aspects are shown separately for the sake of clarity.}
\label{lab:fig_hbetademo}
\end{figure*}
%%%
\begin{figure*}
%\rotate
 \epsfig{figure=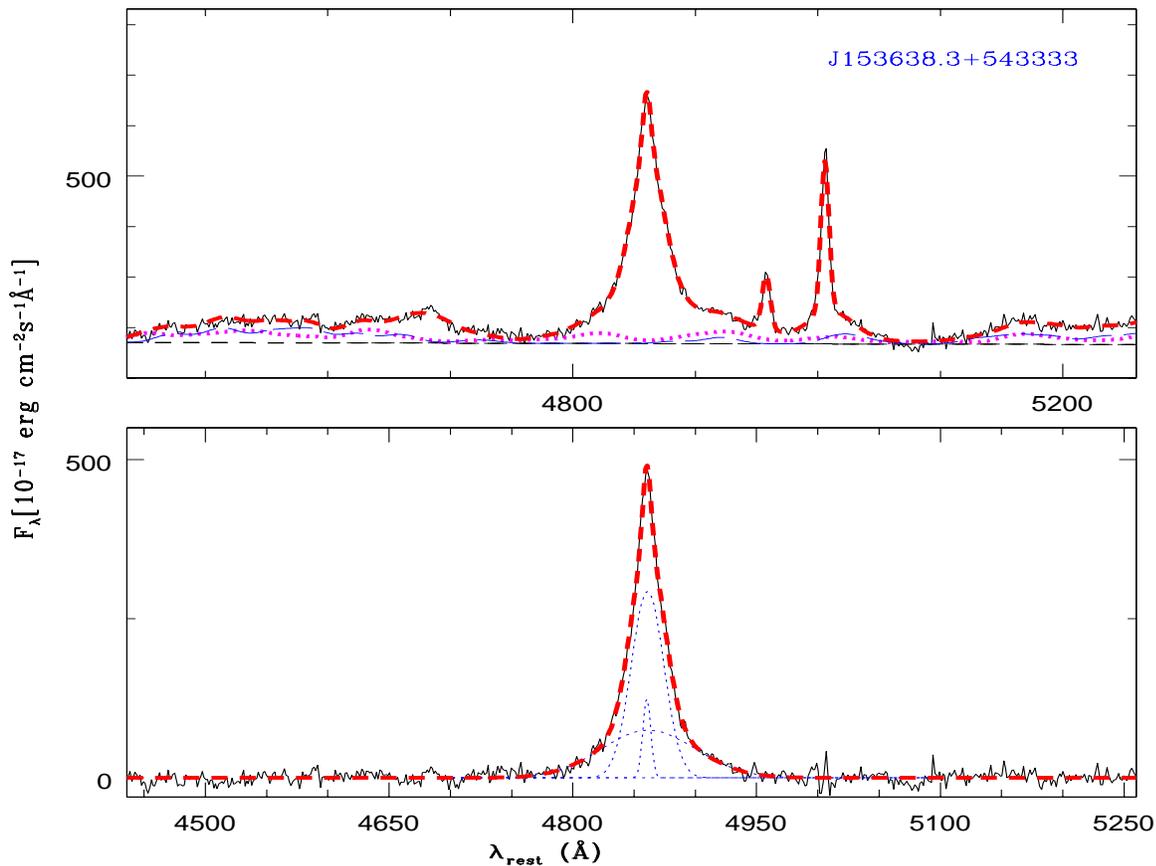,height=12.5cm,width=16.cm,angle=0} 
 \caption{As  in Fig.~\ref{lab:fig_hbetademo} IGO spectra for J153638.3$+$543333.}
\label{lab:fig_hbetademo2}
\end{figure*}

%%%%%%%%%%%%FIGURE 
\begin{figure*}
\epsfig{figure=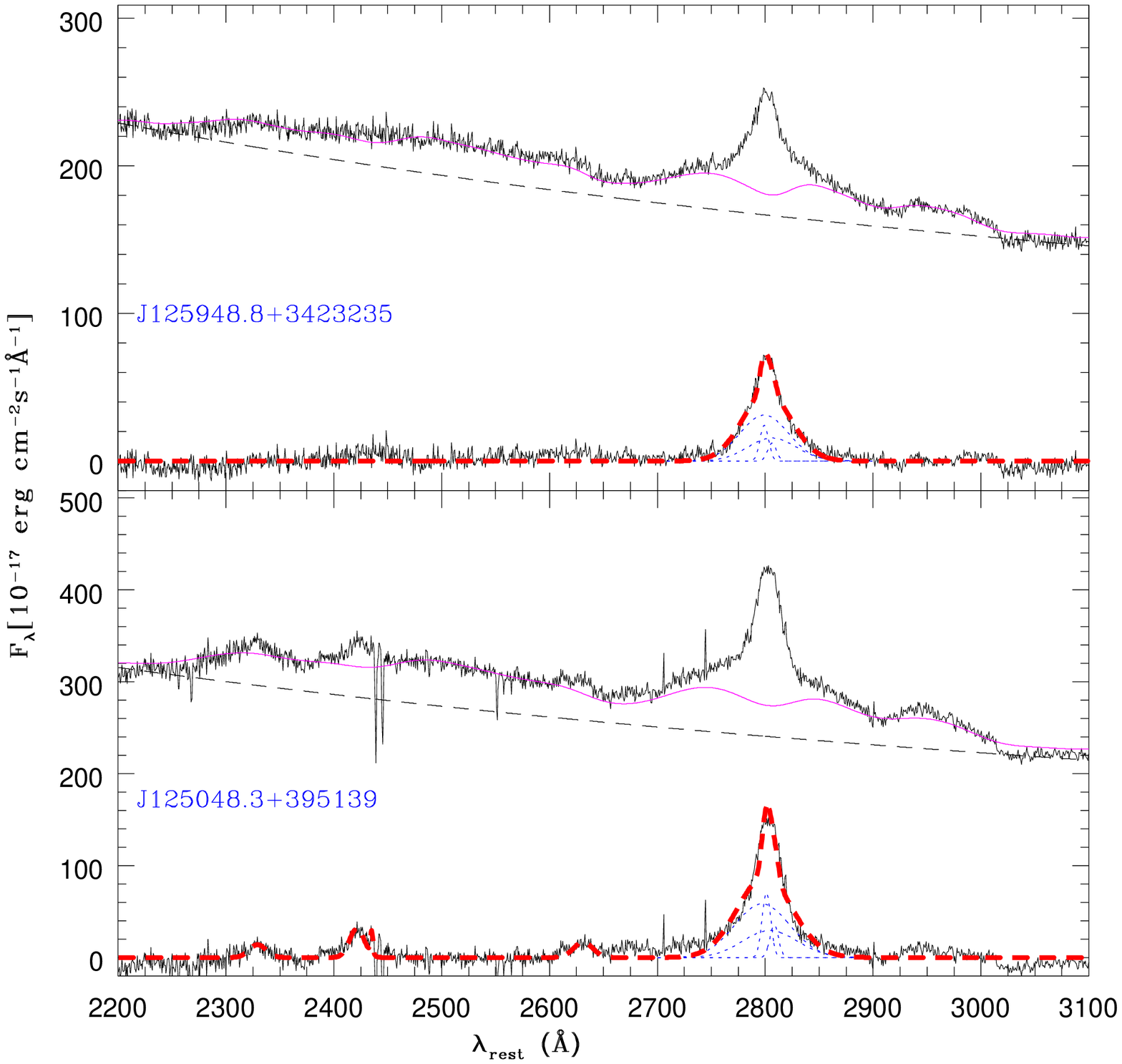,height=12.5cm,width=16.cm,angle=0}
\caption[]{\mgii emission line fits for two RQQSOs, J125948.8+342323.5 and
  J125048.3+395139. The upper plot in each panel shows the SDSS spectra and
  continuum as a dashed line with the best fit UV \feii\ template as a
  solid (magenta) line.  The lower plot in each panel show the
  continuum and \feii\ subtracted spectrum (solid) the best fit
  \mgii profile (thick dashed/red) and \mgiiab\ components by (dotted/blue)
  lines.  Note that the entire fits are performed simultaneously
  (not first continuum subtraction, then \mgii fit) but are shown
  separately for the sake of clarity. }
\label{lab:fig_mgiidemo}
\end{figure*}

\subsection{\hbeta\ region fit}  \noindent
\label{subsec:hbeta}

(i) We fit the spectrum comprising the \hbeta line in the rest
wavelength range between 4435 and 5535~\AA. The continuum in this
region is modeled by a single power law, i.e., $a_{1} \lambda
^{-\alpha}$.\\

(ii) The complex profile of the \hbeta line is fitted with multiple
(one to four) Gaussian with an initial guess of two narrow, one broad
and one very broad components.  To reduce the arbitrariness of the
component fits and also to make the decomposition more physical, we
have constrained the redshift and width of the two narrow components
of \hbeta to be the same as those of the \oiiiab lines. The line
profile of \oiiib (and hence of \oiiia) are modeled as double
Gaussians, with one stronger narrow component with width less than
2000 \kms\ and one weaker broader component with width less than 4000
\kms. However, if for some source spectra the second component is not
statistically required, the procedure we use automatically drops it
during the fit. So, we have only four free parameters, two for
redshift and two for width, of the \oiiiab and narrow
\hbeta\ components.\\

(iii) The optical \feii\ emission is modeled as $C(\lambda) =
c_{b}C_{b}(\lambda) + c_{n}C_{n}(\lambda)$, where $C_{b}(\lambda)$
represents the broad \feii lines and $C_{n}(\lambda)$ the narrow \feii
lines, with the relative intensities fixed at those of I\,ZW\,1, as
given in Tables A.1 and A.2 of V{\'e}ron-Cetty, Joly \& V{\'e}ron
(2004).  The redshift of the broad \feii lines and the
\hbeta\ component is fitted as a free parameter, while their widths
are kept the same, with a constraint that they should be larger than
1000 \kms; this width is then used for estimating the BH masses.
Similarly, for the narrow \feii line the redshift is fitted as a free
parameter but the width is kept the same as that of the stronger
narrow \oiiib component. The fourth (very broad) \hbeta component, if
required for the fit, is subject only to the constraint that its width
should be more than 1000 \kms. The emission lines other than \feii and
\oiiiab (see Table 2 in Vanden Berk et al.\ 2001), are modeled with
single Gaussians.

The final fit is achieved by simultaneously varying all the free
parameters to minimize the \chisq\ value until the reduced $\chi^2_r$
is $\approx 1$.  Samples of our spectral fitting in the optical region
are given for \hbeta in Fig.~\ref{lab:fig_hbetademo} and
Fig.~\ref{lab:fig_hbetademo2}. The values for FWHM and EW (both the
broad component, EW$_B$, and total, EW$_{all}$) for the \hbeta lines
are given in Table~\ref{lab:tabhbeta}. \par

%%%%%%%%%%%%%%%%%%%%%%%%MGII FIT
\subsection{\mgii\ doublet region fit} \noindent
  
(i) We fit the spectrum comprising the \mgii doublet region in the
rest wavelength range between 2200 { and} 3200 \AA. The continuum in
this region is modeled by a single power law, i.e., $a_{1} \lambda
^{-\alpha}$. \\

(ii) For fitting the \mgii\ doublet we have used a Gaussian profile
model (Salviander et al.\ 2007) with the initial guess of two Gaussian
components for each line of the \mgii\ doublet; however, if the second
component is not statistically required the procedure automatically
drops it during the fit. The redshift and width of each component
(narrow/broad) of \mgiia\ were tied to the respective components of
the \mgiib line. 
The peak intensity of \mgiia\ was constrained to be twice that of 
\mgiib\, as is predicted theoretically.

In addition, we have constrained the width of narrow
component to be smaller than 1000 \kms\, and the width of the broader
\mgii component to be same as width of UV \feii\ emission line in the
region (UV \feii). We used an UV \feii template generated by Tsuzuki
et al.\ (2006), basically from the measurements of I\,ZW\,1, which
also employ calculations with the CLOUDY photoionization code (Ferland
et al.\ 1998).  This template is scaled and convolved to the FWHM
value equivalent to the broad components of \mgii\ by taking into
account the FWHM of the I\,ZW\,1 template. The best fit value of the
broad component of \mgii\ obtained in this way is finally used in our
calculation of BH mass.  To test for any overfitting caused by
assuming two components, we also forced our procedure to fit only
single components; however, in doing so for all our sources, a very
good fit is never found for the wings of the lines nor for their
central narrow cores.

(iii) Emission lines other than \feii lines identified from the
composite SDSS QSO spectrum (see Table 2 in Vanden Berk et al.\ 2001),
are modeled with single Gaussians.

The final fit is achieved by varying all the free parameters
simultaneously by minimizing the \chisq\ value, until the reduced
$\chi^2_r$ is $\approx 1$.  Demonstrations of our spectral fitting in
the UV region are given in Fig.~\ref{lab:fig_mgiidemo}, and values for
FWMHs and EWs are given in Table~\ref{lab:tabmgii}.

%% %

\subsection{Optical microvariability and spectral properties}
The results from Paper I, with a modest sample of 37 sources, showed
that the spectral properties for the sources with and without optical
microvariability are quite similar. Here we search for better
statistical results by adding our new 46 sources to the sample in
Paper I.  In these added 46 sources, among the 42 sources for
which we have spectral coverage of the \hbeta line, 9 have shown
optical microvariability while the other 33 have not been seen to show
this property (Table~\ref{lab:tabsamp}). Of the 4 new sources
with spectral coverage of the \mgii doublet, not one showed optical
microvariability. We now investigate the correlation between spectral
properties (i.e., FWHM and EW) and optical microvariability with our
best$-$fitting values and with the larger combined sample of 83
sources. Among the 53 sources with only \hbeta line coverage, optical
microvariability is shown by 15 sources, while of the 22 sources with
only \mgii doublet spectral coverage, only 4 have been shown to
  exhibit microvariability properties. Of the remaining 8 source
  spectra that cover both \mgii and \hbeta lines, optical
microvariability was shown by one source. \par

From the multiple component fit of the \hbeta line we have used the
broad component fit (Sect.\ \ref{subsec:hbeta}) to perform the
comparison. This is because the clouds responsible for this broad
component are clearly in the sphere of influence of the massive BH,
while the other components may not be. In Fig.~\ref{lab:histo_ewfwhm}
we show the histograms of FWHM and rest frame EW values based on our
best fits for \hbeta and \mgii lines. The shaded and non-shaded
regions correspond respectively to sources with and without confirmed
optical microvariability. From these plots it appears that the
distributions of sources with and without microvariability are on the
whole quite similar. To quantify any differences in these
distributions we have performed Kolmogorov-Smirnov (KS) tests on all
the distributions shown in Fig.~\ref{lab:histo_ewfwhm}.  For the null
hypothesis that the samples are drawn from the similar distributions
we found the probabilities for the two $\rm H\beta(\rm EW)$
distributions to be 0.90 and for the two $\rm Mg~II(\rm EW)$
distributions to be 0.70. Similarly, we found the KS-tests null
probability value for the two $\rm H\beta(\rm FWHM)$ distributions to
be 0.29 and for the two $\rm Mg~II(\rm FWHM)$ distributions is 0.99.

%%%%%%%FIGURE
\begin{figure*}
%\rotate
 \epsfig{figure=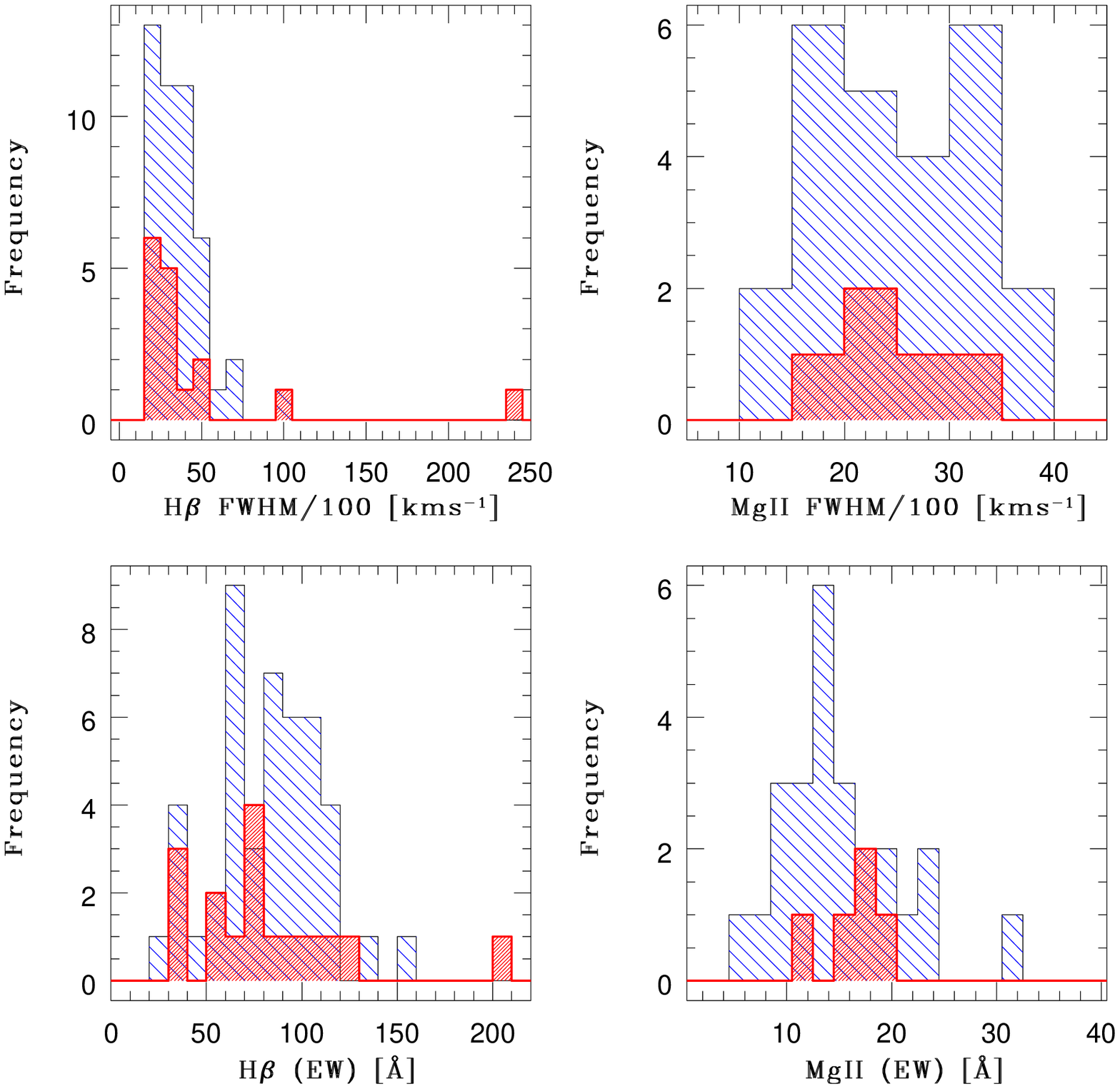,height=12.cm,width=16.cm,angle=0} 
 \caption{Histograms of
FWHM and rest frame equivalent width (EW) based on best fits of
\hbeta and \mgii lines. The red (dark) shaded regions correspond to sources with
confirmed optical microvariability while the striped regions correspond to those sources
for which optical microvariability has not been detected.}
\label{lab:histo_ewfwhm}
\end{figure*}

\begin{figure}
\epsfig{figure=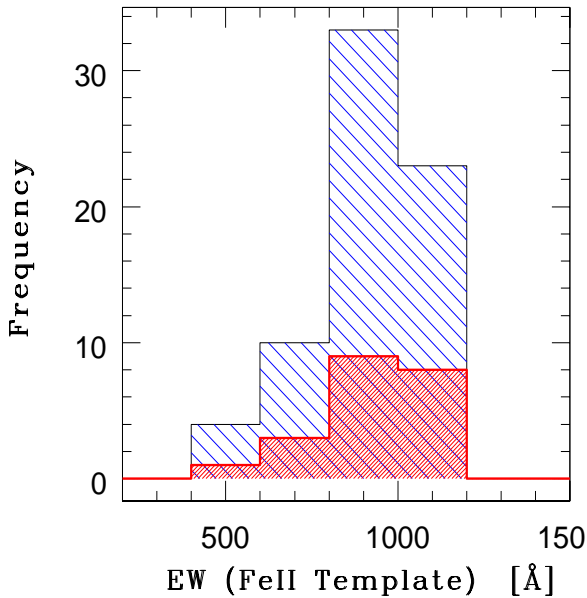,height=8.cm,width=8.cm,angle=00,bbllx=17bp,bblly=435bp,bburx=290bp,bbury=691bp,clip=true}
\caption[]{Histograms of EW of broad \feii template
. The red (dark) shaded regions correspond to sources with
confirmed optical microvariability while the striped regions correspond to those sources
for which optical microvariability has not been detected.}
\label{lab:figfeiifrac}
\end{figure}

\begin{figure}
\epsfig{figure=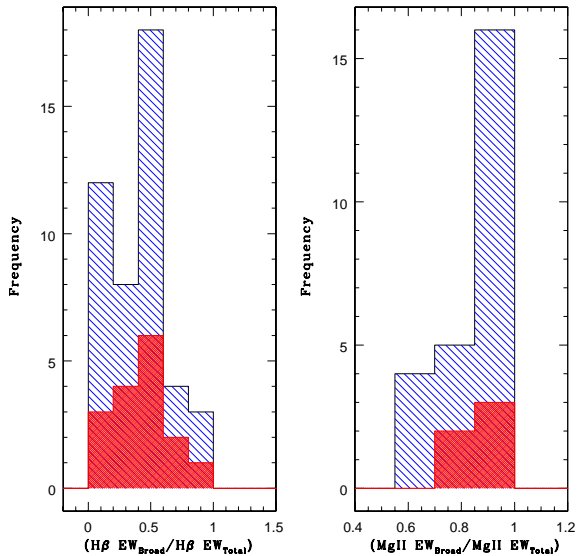,height=8.cm,width=8.cm}
\caption[]{The left panel shows the histograms of the
fraction of the broad  \hbeta  component in the \hbeta emission lines and the right panel shows the 
 broad  \mgii component fraction in \mgii emission lines.}
\label{lab:hbetamgiifrac}
\end{figure}

\begin{figure*}
\epsfig{figure=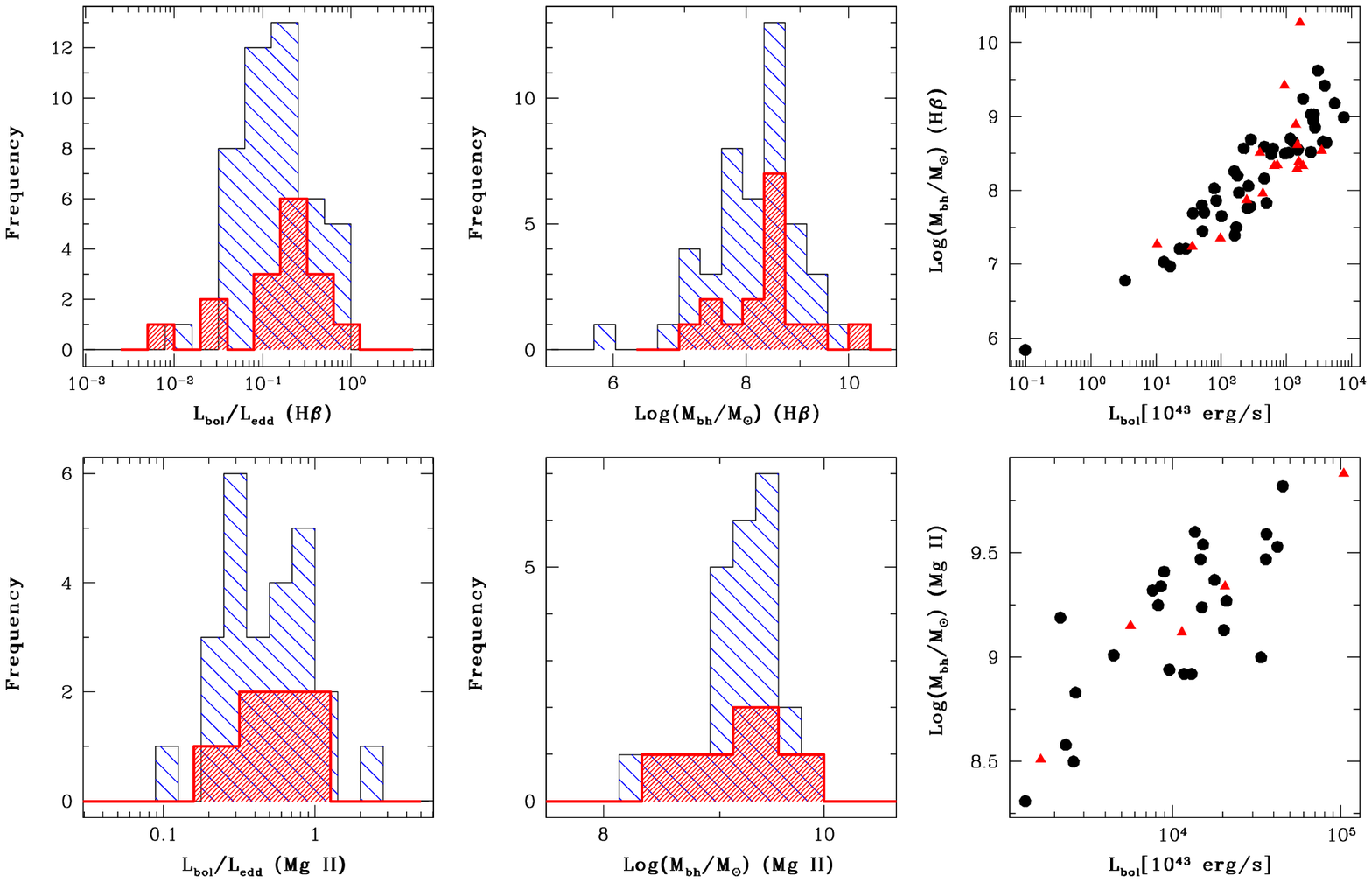,height=10.cm,width=15.cm,angle=00,bbllx=10bp,bblly=338bp,bburx=590bp,bbury=716bp,clip=true}
\caption[]{The upper left and middle panels show the histograms of
Eddington ratio $L_{bol}/L_{edd}$ and ${\rm log}(M_{bh}/M_{\odot})$
based on the \hbeta line, while the corresponding bottom panels show the results
based on the \mgii line. The shaded regions correspond to sources with
confirmed optical microvariability while the non-shaded regions are for
those sources for which optical microvariability is not detected.
The right upper and bottom panels respectively show plots of ${\rm log}(M_{bh}/M_{\odot})$
versus $L_{bol}$.
Triangles and circles respectively show
sources with and without confirmed optical microvariability.}
\label{lab:fig_eddi_ratio}
\end{figure*}

 In addition in the ``blazar component'' scenario one would also
 expect some dilution of emission line strength of the \feii
 template for microvariable sources.  To test this possibility
 we have shown in Fig.~ \ref{lab:figfeiifrac} the EW distribution of
 broad \feii emission template for variable and non-variable sources, which by
 eye appear indistinguishable. The KS-test null probability value for
 these two \feii distributions is 0.72, and so does not give any
 hint of a blazar component.  We have also tried to investigate the
 difference in fraction of broad components of \hbeta and \mgii to the
 respective total line intensity for the sources with and without
 microvariability. We found that the distribution of this fraction of
 the \hbeta and \mgii broad component is quite similar both for
 variable and nonvariable sources as is shown in
 Fig.\ \ref{lab:hbetamgiifrac}}.

The above substantial null-hypothesis probabilities do not
establish any relation of EW or FWHM with the optical microvariability
properties and thus do not support a blazar component model of
microvariability for RQQSOs, which would predict smaller EW values for
variable sources, due to dilution of emission line strength by jet
components (Czerny et al.\ 2008). Here, the results drawn from a
  sample more than twice as large are in agreement with our results
in Paper I. \par

%%%%%%%%%%%%

\begin{table*}
 \centering
 \begin{minipage}{140mm}
  \caption{Results of the \hbeta line analysis}
  \label{lab:tabhbeta}

  \begin{tabular}{@{}llrrrrrrr@{}}

\hline
\multicolumn{1}{c}{QSO Name}  &   \multicolumn{1}{c}{$z_{emi}$ } &
                         \multicolumn{1}{c}{$\frac{L(5100\AA)}{10^{43} \rm erg~s^{-1}}$ }    &
                         \multicolumn{1}{r}{FWHM(km s$^{-1}$)}               &
                         \multicolumn{1}{r}{log($\frac{M_{bh}}{M_{\odot}}$)} &
                         \multicolumn{1}{r}{$\frac{L_{bol}}{L_{edd}}$} &
                         \multicolumn{1}{r}{EW$_B$(\AA)} &
                         \multicolumn{1}{r}{EW$_{all}$(\AA)} \\
 \hline

J000619.5$+$201210 & 0.0262     &    5.16     & 3292.99  &  7.80  &   0.06  &    54.10  &$  113.68   \pm    0.33    $  \\
J002913.6$+$131603 & 0.1462     &  150.94     & 3615.62  &  8.62  &   0.25  &    17.30  &$   36.56   \pm    0.31    $  \\
J004547.3$+$041024 & 0.3825     &  373.16     & 3025.37  &  8.66  &   0.55  &    31.97  &$   65.59   \pm    0.53    $  \\
J005334.9$+$124136 & 0.0604     &   28.55     & 2095.22  &  7.78  &   0.32  &    31.49  &$   58.97   \pm    0.32    $  \\
J005452.1$+$252538 & 0.1651     &   63.57     & 4272.75  &  8.57  &   0.11  &    54.00  &$  119.86   \pm    0.62    $  \\
J011354.5$+$390744 & 0.2325     &   16.18     & 4179.08  &  8.26  &   0.06  &    46.60  &$   84.73   \pm    1.18    $  \\
J012240.6$+$231015 & 0.0528     &    3.77     & 3136.91  &  7.69  &   0.05  &     0.93  &$   37.37   \pm    0.49    $  \\
J051611.4$-$000859 & 0.0334     &    9.91     & 1654.62  &  7.35  &   0.30  &     3.48  &$  118.20   \pm    0.22    $  \\
J051633.4$-$002713 & 0.2922     &  131.20     & 3959.10  &  8.66  &   0.19  &    61.56  &$  105.36   \pm    0.48    $  \\
J055453.6$+$462622 & 0.0245     &    1.35     & 1882.71  &  7.03  &   0.09  &    24.15  &$  158.56   \pm    0.59    $  \\
J071415.1$+$454156 & 0.0560     &    7.97     & 3822.48  &  8.03  &   0.05  &    42.88  &$  115.60   \pm    0.51    $  \\
J073657.0$+$584613 & 0.0392     &    5.62     & 2868.64  &  7.70  &   0.08  &    39.24  &$   94.99   \pm    0.41    $  \\
J084742.4$+$344504 & 0.0640     &   25.27     & 2406.85  &  7.87  &   0.23  &     4.87  &$   66.98   \pm    0.27    $  \\
J092512.9$+$521711 & 0.0355     &    1.06     & 2657.45  &  7.27  &   0.04  &    53.63  &$  121.19   \pm    0.63    $  \\
J092603.3$+$124404 & 0.0289     &    2.34     & 2024.97  &  7.21  &   0.10  &    18.53  &$  102.05   \pm    0.47    $  \\
J095652.4$+$411522 & 0.2334     &  395.74     & 7136.27  &  9.42  &   0.10  &    47.62  &$   87.21   \pm    0.36    $  \\
J101420.7$-$041840 & 0.0582     &    5.26     & 2180.48  &  7.45  &   0.13  &    16.03  &$   35.17   \pm    0.61    $  \\
J105143.9$+$335927 & 0.1671     &   29.00     & 5980.04  &  8.69  &   0.04  &     7.75  &$  102.03   \pm    0.76    $  \\
J110631.8$-$005252 & 0.4232     &  185.83     & 2465.65  &  8.33  &   0.59  &    54.71  &$  100.49   \pm    0.68    $  \\
J111908.7$+$211918 & 0.1759     &   67.27     & 3197.73  &  8.33  &   0.21  &    83.05  &$  201.91   \pm    0.72    $  \\
J112147.1$+$114418 & 0.0500     &   10.33     & 2322.45  &  7.65  &   0.16  &    33.37  &$   66.36   \pm    0.35    $  \\
J112302.3$-$273004 & 0.3902     &  271.41     & 5058.81  &  9.03  &   0.17  &    64.23  &$   67.93   \pm    0.47    $  \\
J112439.2$+$420145 & 0.2243     &   97.42     & 3519.45  &  8.50  &   0.21  &     2.34  &$   87.53   \pm    0.54    $  \\
J115349.3$+$112830 & 0.1763     &   46.83     & 4711.23  &  8.59  &   0.08  &    58.22  &$  105.72   \pm    0.59    $  \\
J120309.6$+$443153 & 0.0017     &    0.01     & 1617.19  &  5.84  &   0.01  &    18.63  &$   28.88   \pm    1.20    $  \\
J121032.6$+$392421 & 0.0032     &    0.34     & 1997.36  &  6.78  &   0.04  &    22.95  &$   81.20   \pm    0.17    $  \\
J132349.5$+$654148 & 0.1677     &   46.77     & 2857.81  &  8.16  &   0.22  &    40.22  &$   73.15   \pm    0.65    $  \\
J140516.2$+$255534 & 0.1651     &   50.57     & 1912.69  &  7.83  &   0.51  &    45.00  &$   73.82   \pm    0.49    $  \\
J141348.3$+$440014 & 0.0897     &   26.76     & 2923.47  &  8.06  &   0.16  &    48.78  &$  139.40   \pm    0.70    $  \\
J141700.7$+$445606 & 0.1475     &   16.46     & 1534.52  &  7.39  &   0.45  &     2.79  &$   35.01   \pm    0.63    $  \\
J142906.6$+$011706 & 0.0866     &    8.55     & 3100.99  &  7.86  &   0.08  &    19.46  &$  108.65   \pm    0.85    $  \\
J144207.4$+$352623 & 0.0772     &   17.34     & 1723.99  &  7.50  &   0.37  &    40.97  &$   62.95   \pm    0.51    $  \\
J153638.3$+$543333 & 0.0387     &    2.93     & 1918.30  &  7.21  &   0.12  &    54.76  &$   99.34   \pm    0.69    $  \\
J155202.3$+$201402 & 0.2518     &   25.79     & 2108.52  &  7.76  &   0.30  &    56.78  &$  111.97   \pm    1.06    $  \\
J162011.3$+$172428 & 0.1145     &   18.14     & 3814.02  &  8.20  &   0.08  &    13.63  &$   52.06   \pm    0.99    $  \\
J170124.8$+$514920 & 0.2914     &  358.12     & 2656.36  &  8.54  &   0.71  &    23.23  &$   35.43   \pm    1.70    $  \\
J175116.6$+$504539 & 0.2974     &  149.87     & 2484.15  &  8.29  &   0.52  &    40.79  &$   78.44   \pm    0.41    $  \\
J211452.6$+$060742 & 0.4850     &  778.92     & 3695.78  &  8.99  &   0.54  &    73.60  &$   81.17   \pm    0.53    $  \\
J213227.8$+$100819 & 0.0629     &   19.10     & 2870.28  &  7.97  &   0.14  &    57.01  &$  104.75   \pm    1.60    $  \\
J221712.2$+$141421 & 0.0655     &   22.58     & 5499.40  &  8.57  &   0.04  &    77.21  &$   93.63   \pm    1.55    $  \\ 
J230315.6$+$085226 & 0.0158     &    3.68     & 1888.66  &  7.24  &   0.14  &    22.62  &$   70.39   \pm    0.84    $  \\
J230702.9$+$043257 & 0.0470     &    1.68     & 1667.88  &  6.97  &   0.12  &     8.19  &$   87.64   \pm    1.31    $  \\

\hline                                                                                
\end{tabular}                                                                         
\end{minipage}                                                                        
\end{table*}

\begin{table*}
 \centering
 \begin{minipage}{140mm}
  \caption{Results of the \mgii line analysis}
  \label{lab:tabmgii}

  \begin{tabular}{@{}lrcclccrr@{}}

\hline
\multicolumn{1}{c}{QSO Name}     & \multicolumn{1}{c}{$z_{emi}$ } &
                          \multicolumn{1}{c}{$\frac{L(3000\AA)}{10^{43} \rm erg~s^{-1}}$ }         &
                         \multicolumn{1}{c}{FWHM(kms$^{-1}$)}              &
                         \multicolumn{1}{c}{Log($\frac{M_{bh}}{M_{\odot}}$)\footnote{Using the McLure \& Dunlop (2004) scaling relation, i.e., Eq.~\ref{logMuv_L30.eq}.}} &
                         \multicolumn{1}{c}{Log($\frac{M_{bh}}{M_{\odot}}$)\footnote{Using the fixed slope of the $r$--$L$ relation from Dietrich et al.\ (2009), their Eq.\ (6).}} &
                         \multicolumn{1}{c}{$\frac{L_{bol}}{L_{edd}}$} &
                         \multicolumn{1}{c}{EW$_B$(\AA)} &
                         \multicolumn{1}{c}{EW$_{all}$(\AA)} \\
 \hline             
 J112731.9$-$304446 & 0.6679    & 5691.58   &1447.36 & 8.53 &  9.00  &2.32  & 13.14 & $13.14 \pm 0.63$\\ 
 J125048.3$+$395139 & 1.0318    & 1294.13   &3029.19 & 8.78 &  9.32  &0.25  & 19.64 & $23.46 \pm 0.31$\\
 J125948.8$+$342323 & 1.3762    & 2577.21   &3297.04 & 9.04 &  9.54  &0.30  & 14.12 & $16.28 \pm 0.27$\\ 
 J135458.7$+$005211 & 1.1253    & 1451.56   &3007.11 & 8.80 &  9.34  &0.27  & 22.30 & $26.21 \pm 0.17$\\
%% \hline 
%%  J001555.1$+$023024 &               &       &          &        &      &       &    &     \\
%%  J012017.2$+$213346 &               &       &          &        &      &       &    &     \\
%%  J220311.5$-$180143 &               &       &          &        &      &       &    &     \\
\hline
\end{tabular}
\end{minipage}
\end{table*}

%%%%%%%%%%%%%%%%%%%%%%%%%%%%%
\begin{table*}
 \centering
 \begin{minipage}{140mm}
  \caption{BH growth time estimates (in units of Gyr)  from the H$\beta$ line analysis}
 \label{lab:grhbeta}

  \begin{tabular}{@{}lc|ccc|rrr|r@{}}

\hline

\multicolumn{1}{c}{object}&
\multicolumn{1}{c}{$z_{emi}$}& 
\multicolumn{6}{c}{Black hole growth time scale}&
\multicolumn{1}{c}{age of the}\\
\multicolumn{1}{c}{ }&
\multicolumn{1}{c}{ }&
\multicolumn{3}{c|}{$L_{bol}$/$L_{edd}$\,=\,1.0}&
\multicolumn{3}{c|}{$L_{bol}$/$L_{edd}$\,obs.\footnote{Values of $\ell$ estimated
                       using H$\beta$ lines (Table~\ref{lab:tabhbeta}).}}& 
\multicolumn{1}{c}{Universe}\\
\multicolumn{1}{c}{ }&
\multicolumn{1}{c}{ }&
\multicolumn{3}{c}{$M_{bh}$(seed)}&
\multicolumn{3}{c}{$M_{bh}$(seed)}&
\multicolumn{1}{c}{(at $z_{emi}$)} \\
\multicolumn{1}{c}{ }&
\multicolumn{1}{c}{ }&
\multicolumn{1}{c}{$10\,M_\odot$}  &
\multicolumn{1}{c}{$10^3\,M_\odot$}&
\multicolumn{1}{c}{$10^5\,M_\odot$}&
\multicolumn{1}{c}{$10\,M_\odot$}  &
\multicolumn{1}{c}{$10^3\,M_\odot$}&
\multicolumn{1}{c}{$10^5\,M_\odot$}&
\multicolumn{1}{c}{{[}$10^9$\,yr]} \\
\hline
J000619.5$+$201210  &  0.0262   &  0.68&   0.48 &  0.28 &   12.40  &   8.75    &   5.11  &   13.11    \\
J002913.6$+$131603  &  0.1462   &  0.76&   0.56 &  0.36 &    3.09  &   2.28    &   1.47  &   11.62    \\
J004547.3$+$041024  &  0.3825   &  0.77&   0.57 &  0.37 &    1.38  &   1.02    &   0.66  &    9.33    \\ 
J005334.9$+$124136  &  0.0604   &  0.68&   0.48 &  0.28 &    2.12  &   1.50    &   0.87  &   12.66    \\ 
J005452.1$+$252538  &  0.1651   &  0.76&   0.56 &  0.36 &    6.60  &   4.86    &   3.11  &   11.41    \\ 
J011354.5$+$390744  &  0.2325   &  0.73&   0.53 &  0.33 &   11.94  &   8.65    &   5.36  &   10.70    \\ 
J012240.6$+$231015  &  0.0528   &  0.67&   0.47 &  0.27 &   12.90  &   9.05    &   5.19  &   12.76    \\
J051611.4$-$000859  &  0.0334   &  0.64&   0.44 &  0.24 &    2.11  &   1.44    &   0.78  &   13.01    \\ 
J051633.4$-$002713  &  0.2922   &  0.77&   0.57 &  0.37 &    4.00  &   2.96    &   1.91  &   10.12    \\ 
J055453.6$+$462622  &  0.0245   &  0.60&   0.40 &  0.20 &    7.03  &   4.70    &   2.37  &   13.13    \\ 
J071415.1$+$454156  &  0.0560   &  0.71&   0.50 &  0.30 &   13.82  &   9.89    &   5.96  &   12.71    \\
J073657.0$+$584613  &  0.0392   &  0.67&   0.47 &  0.27 &    8.84  &   6.20    &   3.56  &   12.93    \\
J084742.4$+$344504  &  0.0640   &  0.69&   0.49 &  0.29 &    3.02  &   2.14    &   1.26  &   12.61    \\ 
J092512.9$+$521711  &  0.0355   &  0.63&   0.43 &  0.23 &   16.55  &  11.27    &   5.99  &   12.98    \\ 
J092603.3$+$124404  &  0.0289   &  0.62&   0.42 &  0.22 &    6.36  &   4.31    &   2.26  &   13.07    \\
J095652.4$+$411522  &  0.2334   &  0.84&   0.64 &  0.44 &    8.20  &   6.25    &   4.30  &   10.69    \\ 
J101420.7$-$041840  &  0.0582   &  0.65&   0.45 &  0.25 &    5.09  &   3.51    &   1.93  &   12.69    \\
J105143.9$+$335927  &  0.1671   &  0.77&   0.57 &  0.37 &   19.28  &  14.27    &   9.25  &   11.39    \\
J110631.8$-$005252  &  0.4232   &  0.74&   0.53 &  0.33 &    1.25  &   0.91    &   0.57  &    9.00   \\
J111908.7$+$211918  &  0.1759   &  0.74&   0.53 &  0.33 &    3.48  &   2.53   &    1.58  &   11.29   \\
J112147.1$+$114418  &  0.0500   &  0.67&   0.47 &  0.27 &    4.25  &   2.97   &    1.69  &   12.79   \\
J112302.3$-$273004  &  0.3902   &  0.81&   0.60 &  0.40 &    4.77  &   3.58   &    2.39  &    9.26   \\
J112439.2$+$420145  &  0.2243   &  0.75&   0.55 &  0.35 &    3.58  &   2.63   &    1.67  &   10.78   \\
J115349.3$+$112830  &  0.1763   &  0.76&   0.56 &  0.36 &    9.40  &   6.92   &    4.44  &   11.29   \\
J120309.6$+$443153  &  0.0017   &  0.49&   0.28 &  0.08 &   48.54  &  28.48   &    8.42  &   13.44   \\
J121032.6$+$392421  &  0.0032   &  0.58&   0.38 &  0.18 &   14.86  &   9.72   &    4.58  &   13.42   \\
J132349.5$+$654148  &  0.1677   &  0.72&   0.52 &  0.32 &    3.26  &   2.35   &    1.44  &   11.38   \\
J140516.2$+$255534  &  0.1651   &  0.68&   0.48 &  0.28 &    1.34  &   0.95   &    0.56  &   11.41   \\
J141348.3$+$440014  &  0.0897   &  0.71&   0.51 &  0.31 &    4.45  &   3.19   &    1.93  &   12.29   \\
J141700.7$+$445606  &  0.1475   &  0.64&   0.44 &  0.24 &    1.41  &   0.97   &    0.53  &   11.61   \\
J142906.6$+$011706  &  0.0866   &  0.69&   0.49 &  0.29 &    8.60  &   6.09   &    3.59  &   12.33   \\
J144207.4$+$352623  &  0.0772   &  0.65&   0.45 &  0.25 &    1.77  &   1.23   &    0.68  &   12.45   \\
J153638.3$+$543333  &  0.0387   &  0.62&   0.42 &  0.22 &    5.10  &   3.46   &    1.82  &   12.94   \\
J155202.3$+$201402  &  0.2518   &  0.68&   0.48 &  0.28 &    2.26  &   1.59   &    0.92  &   10.51   \\
J162011.3$+$172428  &  0.1145   &  0.72&   0.52 &  0.32 &    9.38  &   6.77   &    4.17  &   11.99   \\
J170124.8$+$514920  &  0.2914   &  0.76&   0.56 &  0.36 &    1.07  &   0.79   &    0.50  &   10.13   \\
J175116.6$+$504539  &  0.2974   &  0.73&   0.53 &  0.33 &    1.40  &   1.02   &    0.63  &   10.07   \\
J211452.6$+$060742  &  0.4850   &  0.80&   0.60 &  0.40 &    1.49  &   1.12   &    0.75  &    8.53   \\
J213227.8$+$100819  &  0.0629   &  0.70&   0.50 &  0.30 &    5.03  &   3.59   &    2.14  &   12.63   \\
J221712.2$+$141421  &  0.0655   &  0.76&   0.56 &  0.36 &   18.52  &  13.62   &    8.73  &   12.59   \\
J230315.6$+$085226  &  0.0158   &  0.63&   0.43 &  0.22 &    4.44  &   3.02   &    1.59  &   13.25   \\
J230702.9$+$043257  &  0.0470   &  0.60&   0.40 &  0.20 &    4.87  &   3.24   &    1.61  &   12.83   \\
\hline
\end{tabular}
\end{minipage}
\end{table*}
%%%%%%%%%%%%%%%%%%%%%%
\begin{table*}
 \centering
 \begin{minipage}{140mm}
  \caption{BH growth time estimates (in units of Gyr) from the \mgii line analysis}
   \label{lab:grmgii}

  \begin{tabular}{@{}lc|ccc|ccc|c@{}}

\hline
\multicolumn{1}{c}{object}&
\multicolumn{1}{c}{$z_{emi}$}& 
\multicolumn{6}{c}{Black hole growth time scale}&
\multicolumn{1}{c}{age of the}\\
\multicolumn{1}{c}{ }&
\multicolumn{1}{c}{ }&
\multicolumn{3}{c|}{$L_{bol}$/$L_{edd}$\,=\,1.0}&
\multicolumn{3}{c|}{$L_{bol}$/$L_{edd}$\,obs.\footnote{Values of $\ell$ 
    estimated using  \mgii lines (Table~\ref{lab:tabmgii}).} }&

\multicolumn{1}{c}{Universe}\\
\multicolumn{1}{c}{ }&
\multicolumn{1}{c}{ }&
\multicolumn{3}{c}{$M_{bh}$(seed)}&
\multicolumn{3}{c}{$M_{bh}$(seed)}&
\multicolumn{1}{c}{(at $z_{emi}$)} \\
\multicolumn{1}{c}{ }&
\multicolumn{1}{c}{ }&
\multicolumn{1}{c}{$10\,M_\odot$}  &
\multicolumn{1}{c}{$10^3\,M_\odot$}&
\multicolumn{1}{c}{$10^5\,M_\odot$}&
\multicolumn{1}{c}{$10\,M_\odot$}  &
\multicolumn{1}{c}{$10^3\,M_\odot$}&
\multicolumn{1}{c}{$10^5\,M_\odot$}& 
\multicolumn{1}{c}{{[}$10^9$\,yr]}\\
 \hline
    J112731.9$-$304446        & 0.6679 & 0.80&  0.60 & 0.40 & 0.35  & 0.26  & 0.17  & 7.35\\ 
    J125048.3$+$395139        & 1.0318 & 0.83&  0.63 & 0.43 & 3.34  & 2.54  & 1.73  & 5.63\\ 
    J125948.8$+$342323	      & 1.3762 & 0.86&  0.66 & 0.46 & 2.85  & 2.19  & 1.52  & 4.52\\ 
    J135458.7$+$005211	      & 1.1253 & 0.84&  0.64 & 0.44 & 3.10  & 2.35  & 1.61  & 5.29\\ 
\hline
\end{tabular}
\end{minipage}
\end{table*}

%%%%%%%%%%%%%%%5
%%%%%%%%%%%%
%%%%%%%%%%%%

\begin{figure*}
\epsfig{figure=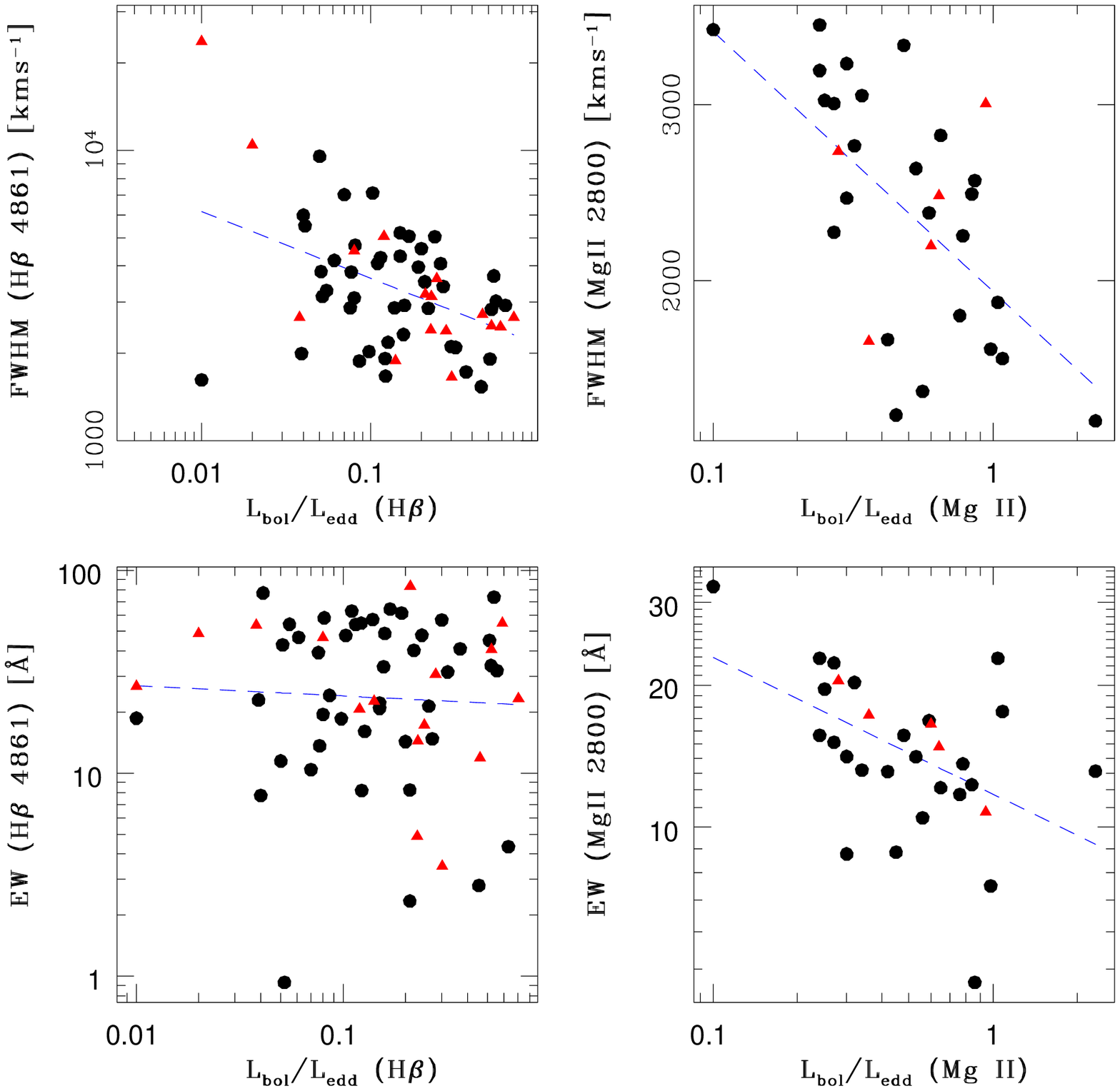,height=12.cm,width=14.cm,angle=0}
\caption[]{Observed variation of FWHM (upper) and EW (lower) with
  Eddington ratio ($\ell={\rm L}_{bol}/{\rm L}_{edd}$) based on both
  \hbeta (left) and \mgii (right) lines.  Triangles and circles
  respectively show sources with and without confirmed optical
  microvariability. { The dashed lines show the linear regression fits
  treating $\ell$ as the independent variable. The
   correlations of both FWHM and EW with $\ell$,  
  suggest that the Eddington ratio is one of the fundamental parameters
  responsible for some AGN properties}.}
\label{lab:lratio_ew_fwhm}
\end{figure*}

\section{Black Hole Mass Measurements, Eddington ratios and black hole growth times}

We have estimated the black hole masses using the virial single epoch
method (Dibai 1980), following the approach in Paper I. This method
has been shown to be consistent with reverberation mapping masses
(e.g., Bochkarev \& Gaskell 2009).  Improved empirical relationships
(e.g., Vestergaard \& Peterson 2006) are used for estimating the
central BH mass based on FWHMs of emission lines such as \hbeta and
Mg~{\sc ii}. We used Eq.~1 and Eq.~2 in Paper I, taken from
Vestergaard \& Peterson (2006) and McLure \& Dunlop (2004), to
determine respectively black hole masses based on \hbeta lines and
\mgii lines, viz;
 \begin{eqnarray}
  \lefteqn{\log \,M_{\rm BH} (\rm H\beta)  =
   \log \,\left[ \left(\frac{\rm FWHM(H\beta)}{1000~km~s^{-1}} \right)^2\right]} \nonumber \\
  & & \mbox{} + (6.91\pm0.02)+ \log \, \left( \frac{\lambda \it L_{\lambda} 
    {\rm (5100\,\AA)}}{10^{44} \rm erg~s^{-1}}\right)^{0.50\pm0.06}, 
\label{logMopt_L51.eq}
  \end{eqnarray}

\begin{eqnarray}
    \lefteqn{\log \,M_{\rm BH} (\rm Mg~II) = (0.62\pm 0.14) \log \left( \frac{\lambda \it L_{\lambda}
        \rm (3000\,\AA)} {10^{44} \rm erg~s^{-1}} \right) }\nonumber \\
    & & \mbox{} + 2 \log \left(\frac{\rm FWHM(Mg~II)}{\rm km s^{-1}} \right) + 0.505 
    \label{logMuv_L30.eq}
  \end{eqnarray}
where $L_{\lambda}{\rm (5100\,\AA)}$ \& $L_{\lambda} {\rm (3000\,\AA)}$
  are the monochromatic luminosity at 5100\,\AA~ and 3000\,\AA~, 
  respectively, which we have computed from the best fit power-law
  continuum, $a_{1} \lambda ^{-\alpha}$, in our simultaneous fit of
  the whole spectral region.

We have also estimated the Eddington ratio $\ell \equiv
L_{bol}/L_{edd}$, where $L_{bol}$ is taken as $ 5.9 \times \lambda
L_{\lambda}$(3000\AA) and $9.8 \times \lambda L_{\lambda}$(5100\AA)
for \mgii and \hbeta, respectively (McLure \& Dunlop 2004), and
$L_{edd}=1.45 \times 10^{38} (M_{bh}/M_{\odot}) \ \ \rm erg~s^{-1} $,
assuming a mixture of hydrogen and helium so the mean molecular weight
is $\mu=1.15$.  The combined sample results are given in
Fig.\,\ref{lab:fig_eddi_ratio}, which by eye indicates that
distributions of sources with and without optical microvariability
appear similar with respect to both BH mass and
$\ell$. Quantitatively, this is also supported by KS tests, which show
that the probability of the null hypothesis ($P_{null}$) for sources
with and without microvariation, is as high as 0.69 for BH mass
distributions and 0.11 for Eddington ratio ($\ell$) distributions
based on \hbeta lines fit. Similarly using \mgii lines, the null hypothesis 
($P_{null}$) for sources with and without microvariation, is as high as 0.98 for BH mass
distributions and 0.94 for Eddington ratio ($\ell$) distributions.

To test the reasonableness of estimated Eddington ratios, we also
computed black hole growth times to compare them with the age of the
Universe (at the time the QSO is observed), by using the following
equation (Dietrich et al.\ 2009),

\begin{equation}
  M_{bh}(t_{obs}) = M_{bh}^{seed}(t_0) \,
   {\rm exp}\biggl(\ell \,{(1-\epsilon) \over \epsilon}\,{\tau \over  t_{edd}}\biggr)~,
\label{growthtau.eq}
\end{equation}
\noindent
where $\tau = t_{obs} - t_0$ is the time elapsed since the initial
time, $t_0$, to the observed time, $t_{obs}$; $M_{bh}^{seed}$ is the
seed BH mass; $\epsilon$ is the efficiency of converting mass to
energy in the accretion flow, and $t_{edd}$ is the Eddington time
scale, with $t_{edd} =\sigma_T c / 4 \pi G m_p = 3.92 \times 10^8$\,yr
(Rees 1984). We used Eq.~\ref{growthtau.eq} to derive the times,
$\tau$, necessary to accumulate the BH masses listed in
Tables~\ref{lab:tabhbeta} and \ref{lab:tabmgii}, for seed black holes
with masses of $M_{bh}^{seed} = 10 M_\odot$, $10^3 M_\odot$, and $10^5
M_\odot$, respectively. Two cases are considered: (i) BHs are
accreting at the Eddington-limit, i.e., $\ell
$\,=\,L$_{bol}$/L$_{edd}$\,=\,1.0 and the efficiency of converting
mass into energy is $\epsilon=0.1$; (ii) BHs are accreting with our
observed Eddington ratios and $\epsilon=0.1$.  These results are
summarized in Tables~\ref{lab:grhbeta} and \ref{lab:grmgii}, showing
that the key criterion that black hole growth time should be smaller
than the age of Universe is fulfilled for all sources.

\begin{figure*}
\epsfig{figure=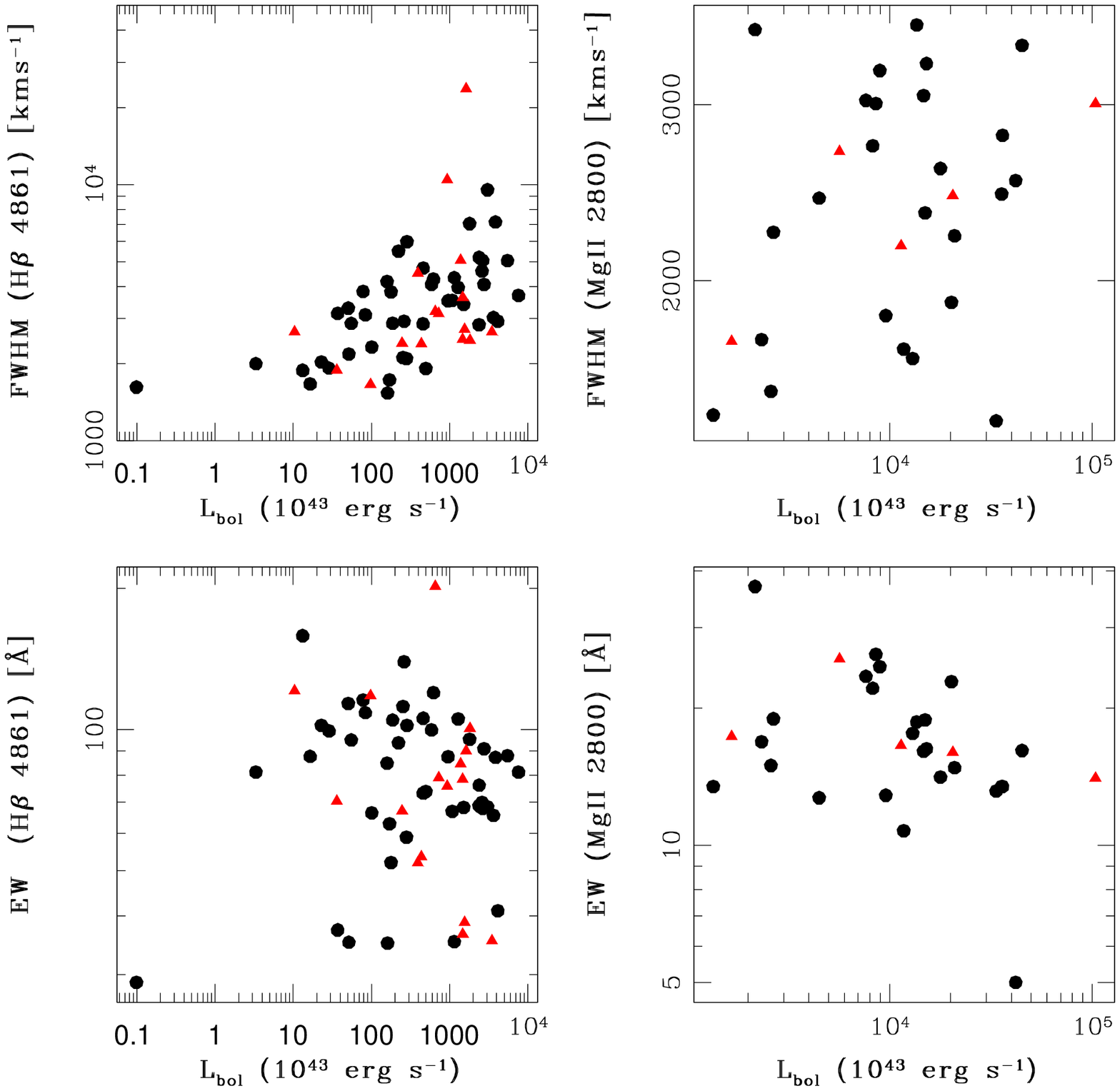,height=12.cm,width=14.cm,angle=0}
\caption[]{As in Fig.~\ref{lab:lratio_ew_fwhm} with $L_{\rm bol}$ as the
independent variable. This panel indicates that the sources with and without optical microvariability
are  similar in  luminosity.}
\label{lab:lbolo_ew_fwhm}
\end{figure*}

In our Paper I we found a weak negative correlation between \hbeta EW
and the Eddington ratio, $\ell$, and a significant one between the
\mgii EW and $\ell$.  In Fig.~\ref{lab:lratio_ew_fwhm} we show the
observed variations of FWHM and EW with $\ell$ based on both \hbeta
and \mgii lines for our combined sample. There appear to be linear
relations of both FWHM and EW with $\ell$ in these log-log plots.  For
the FWHM plots this is unsurprising since the BH masses are
proportional to $L_{edd}$. To quantify any such linear relations for
the EW plots we perform linear regressions, treating $\ell$ as the
independent variable, and find
\begin{eqnarray}
 \rm logEW(H\beta)=(1.33\pm0.12)+(-0.05\pm0.13)\rm  log (\ell) , \nonumber \\
 \rm logEW(\rm Mg~II) =(1.07\pm0.04)+(-0.29\pm0.10)\rm  log (\ell) .
\label{regr_hbeta.eq}
\end{eqnarray}
Here the errors on the fit parameters are purely statistical.  We have
calculated the Spearman rank correlations of logEW with log$\ell$, and
found the correlation coefficient for r$_{s}(H\beta)=-0.05$, with null
probability $p_{null}=0.72$, and so no significant correlation is present. Whereas r$_{s}(\rm
Mg~II)=-0.45$, with $p_{null}=0.013$ and so this negative correlation
is significant. These results are in agreement with those from our
Paper I that were based on a modest sample size.
\section{Discussion and Conclusions}

Modern surveys such as SDSS have allowed investigators to carry out
spectral analyses of large numbers of quasars to estimate central BH
masses using virial approaches (e.g., Shen et al.\ 2008; Fine et
al.\ 2008; Vestergaard \& Osmer 2009) and to understand their
demography (Dong et al.\ 2009b).  These studies have resulted in very
important insights on the important physical parameters of AGN central
engines and their environments.\par

Dong et al.\ (2009b) found that the variation of the emission-line
strength in AGNs are regulated by $\ell$, presumably because it
governs the global distributions of the properties such as column
density of the clouds gravitationally bound in line emitting region.
Shen et al.\ (2008) have found that the line widths of \hbeta and
\mgii follow lognormal distributions with very weak dependencies on
redshift and luminosity. Fine et al.\ (2008) used \mgii lines to
estimate BH masses, and found that the scatter in measured BH masses
is luminosity dependent, showing less scatter for more luminous
objects.

The sample we have considered more than doubles our sample in Paper I,
but still is much smaller compared to those in the above papers, that
consist of between 1100 and almost 57,700 quasars.  However, each
member in our sample has been carefully selected to be among the
special group of RQQSOs and Seyfert galaxies already examined for
optical microvariability (e.g., Carini et al.\ 2007).  This criterion
demands that modest aperture (usually 1--2 m) telescopes can make
precise photometric measurements in just a few minutes, and so limits
the members to the rare QSOs with bright apparent magnitudes (usually
$m_V < 17.5$).  In addition, we have taken special care in the
fitting of the line profiles as discussed in Sections 4.1 and 4.2.\par

It has been found that for BL Lacertae objects the emission line
detection and strength varies with overall continuum flux (e.g.,
Nilsson et al.\ 2008). For instance, when BL Lacs are in optically
faint states weak emission lines have sometimes been detected that are
usually not seen during their high states, presumably due to their
being swamped by the Doppler boosted continuum arising in a strong jet
component (Nilsson et al.\ 2009).

To check whether the continuum states of our objects, high (bright) or
low (faint), have any such effect on their spectral properties, we
have searched in the SDSS DR8 archive for multi-epoch photometric and
spectroscopic observations for the 44 SDSS sources in our sample.  We
found multi-epoch photometric fluxes were available for 14 sources,
with time gaps ranging from a few days to years. We then searched for
spectroscopic observations of these sources and found multi-epoch
spectra are available only for three sources, viz.
J025937.46$+$003736.3; J084030.0$+$465113; J093502.6$+$433111. Among
them, J093502.6$+$433111 has two spectra with a roughly one year time
gap (observations on 2002 Feb 20 and 2003 Jan 31) but only a single
photometric data point at all close to either of those dates (2001 Dec
20). For J084030.0$+$465113 two spectra were available about a month
apart (2001 Jan 13 and 2001 Feb 19) but these had their two closest
photometric observations made on 2001 Jan 24 and 2001 Jan 26,
respectively. As these two photometric observations have only a two
day time gap and have a g-mag difference of only 0.01 this object also
does not serve to test for such a correlation.  We are left with just
one source, J025937.46$+$003736.3, that has both photometric and
spectroscopic multi-epoch data that are well paired. For this source
there were two spectral observations on 2000 Nov 25 and 2001 Sep 21
and corresponding photometric observations on 2000 Nov 27 and 2001 Sep
21 with g$-$mags of 16.41 and 16.47, respectively. As this g$-$mag
difference of 0.06 is much larger than the typical RQQSOs INOV
magnitude variation of about 0.01 mag over a night, this source is the
only one that might test whether brighter or fainter states have any
significant effect on spectral properties.  Even though this
difference in brightness is modest, we carried out our spectral fits
to each of those two spectra to estimate the FWHMs and EWs of the
\hbeta line, using the same method as we used for other sources
(Sect. 4.1). The best fit EW(\hbeta) of broad component for first
epoch was found to be $30.47\pm0.63$\AA \ and for second epoch,
$29.23\pm0.45$\AA.  These values are statistically indistinguishable,
so there is no evidence for any effect due to variation in source
brightness.  However, in order to say anything firm about this
possibility for RQQSOs, nearly simultaneous spectroscopic and
photometric measurements would need to be made on at least two
occasions for a decent sized sample.
 
In this paper we continued our work begun in Paper I with 37
  RQQSOs by analyzing spectra for an additional 46 sources. Most of
these spectra were obtained by us at the IGO.  We conclude that there
is a significant negative correlation between \mgii EW and $\ell$,
(Fig.~\ref{lab:lratio_ew_fwhm}; Section~5), as also found in Paper I
(there Fig.~7) and by Dong et al.\ (2009b).  However, we have not
found any significant correlation between the equivalent widths
  of the \hbeta lines and the Eddington ratio, $\ell$. We can also
see from Fig.~\ref{lab:lratio_ew_fwhm} that there is a decline in FWHM
with $\ell$; this is not surprising since the BH masses are
proportional to $L_{edd}$.  In Paper I (Fig.~8), we noticed a
interesting tendency for sources with detectable optical
microvariability to have somewhat lower luminosity than those with no
such detections, which required investigation with a larger sample.
One might expect such a trend, as lower mass BHs would have
correspondingly shorter physical timescales.  However, as can be seen
from Fig.~\ref{lab:lbolo_ew_fwhm} there is no such trend seen with our
new larger sample of microvariable sources (16 as compared to 7 in
Paper I).

We also find that the BH masses estimated from the FWHMs of both the
\hbeta and \mgii lines are reasonable, in that growth to their
estimated masses from even small seed BHs are easily possible within
the age of the Universe at their observed redshift if the mean $\ell$
values are close to unity (Tables 5 and 6).  This remains true for the
great majority of RQQSOs  even if the value of $\ell$ we compute
from the current continuum flux was constant until the time we observe
them; however, this assumption does not work for 8 out of the 42 QSOs
with \hbeta lines, while we consider $M_{bh}(seed)=10M_{\odot}$, for 4
while we consider $M_{bh}(seed)=10^3M_{\odot}$ and for none while we
consider $M_{bh}(seed)=10^5M_{\odot}$, suggesting that in a few cases
$M_{bh}(seed)$ could need to be as large as $10^5M_{\odot}$ or
  the accretion rate was substantially higher in the past.  With
\mgii lines profiles no QSO was found problematic with respect
  to this assumption (Table~6).

As Fig.~\ref{lab:histo_ewfwhm} shows, histograms of FWHM and rest
frame EW values for sources with and without confirmed optical
microvariability are on the whole quite similar. Under the null
hypothesis that the samples are drawn from the similar distribution
using KS-tests we find a probability of 0.90 for the $\rm H\beta(\rm
EW)$ distributions and 0.70 for the $\rm Mg~II(\rm EW)$ distributions.
Similarly this null probability for the FWHM distributions is 0.29 for
$\rm H\beta(\rm FWHM)$ and 0.99 for $\rm Mg~II(\rm FWHM)$
distributions.  We conclude that EW or FWHM distributions (for both
the \hbeta or \mgii) are probably independent of the presence or
absence of detected microvariability properties in those QSOs and
Seyferts.  As discussed in the Introduction and in more detail in
Paper I, if much of the optical emission in RLQSOs comes from a jet,
then we would expect the EWs of the RLQSOs to be significantly lower
than those of the RQQSOs and that the EWs of microvariable sources
would be less than those of non-variable sources.  Our results are in
agreement to the conclusion in Paper I and thus do not support the
hypothesis (e.g., Gopal-Krishna et al.\ 2003; Czerny et al.\ 2008)
that RQQSOs possess jets that are producing rapid variations.  Instead
it may indicate that variations involving the accretion disc (e.g.,
Wiita 2006) play an important role here.

Further improvements to our results could be obtained through
extensive searches for INOV in a larger sample of RQQSOs to reduce the
statistical uncertainties. In any such studies it would be most useful
to take the spectra just before or after the photometric monitoring
run, so as to ensure the simultaneity of light curve and spectral
properties.  This would rule out any possibility of change in either
of these properties due to temporal gaps between the spectral and
photometric epochs; while we very much doubt that this is an important
effect, it could have an impact on our results.  Such larger samples
could be optimally designed if they were made as homogeneous as
possible on the basis of apparent magnitudes, redshifts and absolute
magnitudes.

\section*{Acknowledgments}

We gratefully acknowledged the observing help rendered by
  Dr. Vijay Mohan and the observing staff at IGO-2m telescope. \par

 Funding for the SDSS and SDSS-II has been provided by
the Alfred P. Sloan Foundation, the Participating Institutions, the
National Science Foundation, the U.S. Department of Energy, the
National Aeronautics and Space Administration, the Japanese
Monbukagakusho, the Max Planck Society, and the Higher Education
Funding Council for England. The SDSS Web Site is
http://www.sdss.org/.
    The SDSS is managed by the Astrophysical Research Consortium for
    the Participating Institutions. The Participating Institutions are
    the American Museum of Natural History, Astrophysical Institute
    Potsdam, University of Basel, University of Cambridge, Case
    Western Reserve University, University of Chicago, Drexel
    University, Fermilab, the Institute for Advanced Study, the Japan
    Participation Group, Johns Hopkins University, the Joint Institute
    for Nuclear Astrophysics, the Kavli Institute for Particle
    Astrophysics and Cosmology, the Korean Scientist Group, the
    Chinese Academy of Sciences (LAMOST), Los Alamos National
    Laboratory, the Max-Planck-Institute for Astronomy (MPIA), the
    Max-Planck-Institute for Astrophysics (MPA), New Mexico State
    University, Ohio State University, University of Pittsburgh,
    University of Portsmouth, Princeton University, the United States
    Naval Observatory, and the University of Washington.   \par
This research has made use of the NASA/IPAC Extragalactic Database
    (NED) which is operated by the Jet Propulsion Laboratory,
    California Institute of Technology, under contract with the
    National Aeronautics and Space Administration.

 \label{lastpage}
%% %%\bibliographystyle{alpha} %% give your .bst file
%% %%\bibliography{ms_joshi}   %% give your .bib file
\end{document}